\documentclass[]{aastex}

\usepackage{epsf}
\usepackage{graphics}
\usepackage{emulateapj5}

\newcommand{\kms}{km\,$\,{\rm s}^{-1}$}
\newcommand{\kmsmpc}{km\,$\,{\rm s}^{-1}\,{\rm Mpc}^{-1}$}
\newcommand{\dv}{$R^{1/4}\,$}

\newenvironment{inlinefigure}{
\def\@captype{figure}
\noindent\begin{minipage}{0.999\linewidth}\begin{center}}
{\end{center}\end{minipage}\smallskip}

\newenvironment{inlinetable}{
\def\@captype{table}
\noindent\begin{minipage}{0.999\linewidth}\begin{center}}
{\end{center}\end{minipage}\smallskip}

\medskip

\slugcomment{Submitted to ApJ}

\shorttitle{} \shortauthors{Koopmans et al.}

\received{}
\begin{document}

\title{The Hubble Constant from the Gravitational Lens B1608+656$^1$}
\footnotetext[1]{Based on observations collected at W.~M. Keck
Observatory, which is operated jointly by the California Institute of
Technology and the University of California, and with the NASA/ESA
Hubble Space Telescope, obtained at STScI, which is operated by AURA,
under NASA contract NAS5-26555.}

\author{L.V.E.\ Koopmans$^{2,4}$, T.Treu$^3$, C.D. Fassnacht$^5$,
R.D. Blandford$^2$, G. Surpi$^{2,6}$}

\footnotetext[2]{California Institute of Technology, Theoretical
Astrophysics, mailcode 130-33, Pasadena, CA 91125}

\footnotetext[3]{California Institute of
Technology, Astronomy, mailcode 105-24, Pasadena, CA 91125}

\footnotetext[4]{Current Address: STScI, 3700 San Martin Drive,
Baltimore MD 21218}  

\footnotetext[5]{Department of Physics, University of California,
1 Shields Ave., Davis, CA  95616}  

\footnotetext[6]{Current Address: FairIsaac, 5935 Cornerstone
Ct. West, San Diego}

\medskip

\begin{abstract}

We present a refined gravitational lens model of the four-image lens
system B1608+656 based on new and improved observational constraints:
(i) the three independent time-delays and flux-ratios from Very Large
Array (VLA) observations, (ii) the radio-image positions from Very
Large Baseline Array (VLBA) observations, (iii) the shape of the
deconvolved Einstein Ring from optical and infrared {\em Hubble Space
Telescope} (HST) images, (iv) the extinction-corrected lens-galaxy
centroids and structural parameters, and (v) a stellar velocity
dispersion, $\sigma_{\rm ap}=247\pm35$~\kms, of the primary lens
galaxy (G1), obtained from an echelle spectrum taken with the Keck--II
telescope. The lens mass model consists of two elliptical mass
distributions with power-law density profiles and an external shear,
totaling 22 free parameters, including the density slopes which are
the key parameters to determine the value of H$_0$ from lens time
delays.  This has required the development of a new lens code that is
highly optimized for speed. The minimum--$\chi^2$ model reproduces all
observations very well, including the stellar velocity dispersion and
the shape of the Einstein Ring. A combined gravitational-lens and
stellar dynamical analysis leads to a value of the Hubble Constant of
H$_0$=75$^{+7}_{-6}$~\kmsmpc\ (68\% CL; $\Omega_{\rm m}$=0.3,
$\Omega_{\Lambda}$=0.7). The non-linear error analysis includes
correlations between all free parameters, in particular the density
slopes of G1 and G2, yielding an accurate determination of the random
error on H$_0$. The lens galaxy G1 is $\sim$5 times more massive than
the secondary lens galaxy (G2), and has a mass density slope of
$\gamma_{\rm G1}'=2.03^{+0.14}_{-0.14} \pm 0.03$ (68\% CL) for
$\rho\propto r^{-\gamma'}$, very close to isothermal
($\gamma'$=2). After extinction-correction, G1 exhibits a smooth
surface brightness distribution with an \dv\ profile and no apparent
evidence for tidal disruption by interactions with G2. Given the scope
of the observational constraints and the gravitational-lens models, as
well as the careful corrections to the data, we believe this value of
H$_0$ to be little affected by known systematic errors
($\la$\,5\%). 
\end{abstract}


\section{Introduction}

The physics behind gravitational lensing is well understood and
primarily based on gravity, i.e., General Relativity
\citep[e.g.,][]{SEF}. Since \citet{refsdal} it has been known that
arcsecond-scale strong gravitational lenses -- those with
multiply-imaged sources -- provide a tool to measure the expansion
rate of the Universe, i.e., the Hubble Constant (H$_0$), if the mass
distribution of the lens(es) and the time-delay(s) between the lensed
images are known. This provides an elegant one-step and global method
to measure H$_0$, independent of local distance-ladder techniques
\citep[e.g., ][]{parodi00,saha01,keyproject}, which could be prone to
the accumulation of unknown systematics, and seem difficult to improve
beyond the current 10\% precision on the value of H$_0$.

\begin{figure*}[t]
\begin{center}
  \leavevmode
\hbox{%
  \epsfxsize=15.4cm
  \epsffile{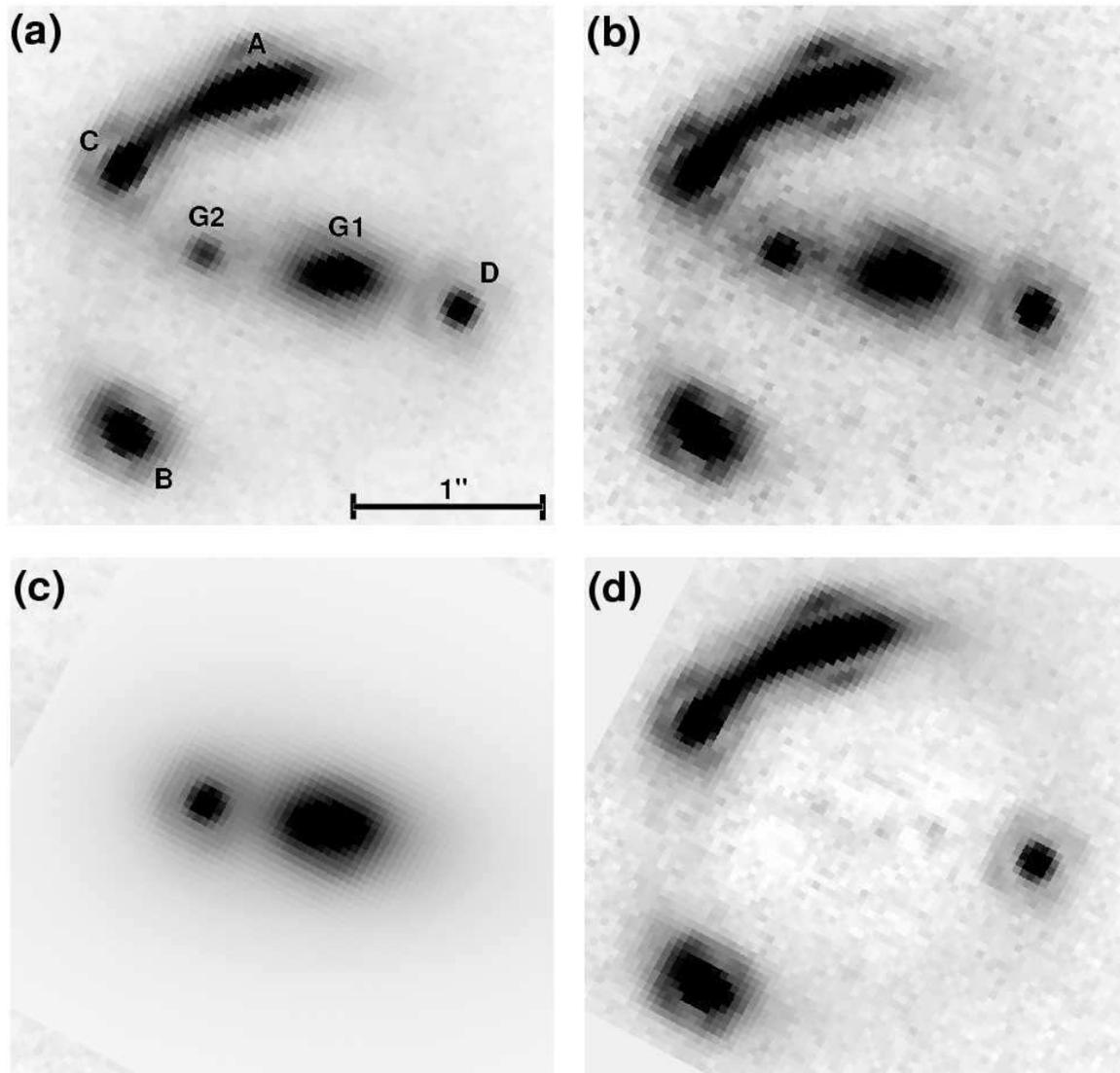}}
\end{center}
\caption{\label{fig:HST} A compilation of the HST--F160W images of
B1608+656: (a) The original reduced image of B1608+656. (b) The
extinction-corrected image (see text). (c) The model surface
brightness distributions of G1 and G2, fitted to the
extinction-corrected brightness distributions. (d) The original image
of B1608+656 after subtraction of the lens galaxy models extincted by
the inferred dust-screen.}
\end{figure*}

More recently, CMB observations with WMAP have been combined with
different cosmological data sets (i.e., Ly-$\alpha$ and Large Scale
Structure observations) to indirectly infer a value of
H$_0$=71$^{+4}_{-3}$~\kmsmpc\ \citep{bennett03,spergel03} similar to
that from the {\em Hubble Space Telescope} (HST) Key-Project,
who find H$_0$=72$^{+8}_{-8}$~\kmsmpc\
\citep{keyproject}. Because H$_0$ can not be measured directly from
the CMB data alone, a combination with other datasets is necessary,
and thus it is prone to its own systematic errors
\citep[e.g.][]{seljak03, bridle03}. Even so, the agreement between the
different values of H$_0$ from different techniques, including those
from Sunyaev-Zeldovich observations \citep[e.g.,][]{szh0}, is
striking. Since H$_0$ is one of the most fundamental cosmological
parameters, however, its measurement should be subject to multiple
cross-examinations, with the ultimate goal of breaking the 10\%
precision limit.

Although gravitational lensing can provide a straightforward and
potentially very precise measurement of H$_0$, its major problems are
that it requires (i) accurately measured time-delays obtained from
long monitoring campaigns and (ii) an accurate determination of the
mass distribution in the lenses and the field (i.e., the ``lens
potential''). At present about a dozen lens systems have measured
time-delays with different degrees of precision \citep[see][for a
recent summary]{courbin02}. This has shifted attention to
understanding the often underestimated complexity of the mass
distribution of the lens. For example, the first discovered
gravitational lens system Q0957+561 \citep{0957discovery}, has an
exquisitely measured time-delay with an error $\la$1\%
\citep[e.g.][]{tk0957_2, 1997ApJ...479L..89O, 1999ApJ...510...64H,
2000ApJ...540..104C, 2001ApJ...552...81O, 2001Ap&SS.275..385S,
2002MNRAS.334..905G, 2003A&A...402..891O}, but a complex lens
potential that includes a nearby cluster. Even though much effort has
been put into constraining this particular lens system, the lens
potential is still not known to adequate precision to allow a
satisfactory determination of~H$_0$ \citep[][and references therein]
{keeton0957}.

One of the dominant degeneracies in gravitational lens models -- even
for systems that are relatively isolated and have no major external
perturbers (e.g., groups or clusters) -- is that between the radial
mass profile and the inferred value of H$_0$. Often point-image
constraints (i.e., positions and flux-ratios) can be satisfied equally
well by mass distributions that have very different slopes of their
radial mass profile \citep[e.g.,][]{1994AJ....108.1156W,
1994RPPh...57..117R, 1995ApJ...443...18W, 2000ApJ...544...98W,
2001glrp.conf..157W, 2001glrp.conf..103S, wucknitz02, nonparam}. This
is similar to the well-known mass-sheet degeneracy
\citep{masssheet}. In general, steeper (shallower) mass profiles lead
to higher (lower) inferred values of H$_0$. To break this degeneracy
and reliably measure H$_0$, one has to determine the mass profile of
the lens galaxies. Often the most effective way is to use additional
external constraints, for example from stellar kinematics of the lens
galaxy \citep[e.g.,][]{TK02a,KT02}.

An often cited result is that of the gravitational lens system
PG1115+080, which gives a value of H$_0$=44$\pm$4~\kmsmpc\ (Impey et
al.~1998) for an isothermal lens mass distribution ($\rho\propto
r^{-2}$). Not only is this value low compared to most other lens and
non-lens measurements, but the quoted error does not include one of
the major sources of uncertainty -- the unknown radial mass
profile. In fact, early-type galaxies can have a considerable scatter
in their radial mass profiles (e.g., Gerhard et al.\ 2001) around the
effective radius, and steeper mass density profiles can increase H$_0$
well above 60 \kmsmpc\, (Impey et al.\ 1998).

It was recently shown that, if one includes the observed stellar
velocity dispersion \citep{tonry98} of the lens galaxy in PG1115+080
as additional constraint, the effective slope of the mass density
profile can be measured directly and also properly included in the
error estimate on H$_0$. The effective slope of the density profile
for this particular system is found to be steeper than isothermal,
leading to a higher value of H$_0$=59$^{+12}_{-7} \pm 3$~\kmsmpc\
\citep{TK02b}. This measurement is in better agreement with most other
methods and its larger error bars include the contribution of the
residual uncertainty on the mass distribution of the lens. This
analysis has led us to initiate a new program with Keck and the {\sl
Very Large Telescope} (VLT) to measure the stellar velocity dispersion
of a number of additional systems with known time-delays, in order to
determine a more precise value of H$_0$ from gravitational lensing,
less affected by the radial mass profile degeneracy.

In this paper, we focus on the gravitational lens system B1608+656
(source redshift $z_s=1.39$; lens redshift $z_l=0.63$)
\citep{sm1608,1995ApJ...447L...9S,zs1608}.  The system is unique in
that all three time delays are known between its four lensed images
\citep{F99,F02}, with accuracies of a few percent. In addition,
multi-color images are available in the HST archive that show the
lensed arcs of the radio-source host galaxy connecting to a full
Einstein Ring \citep[e.g.,][]{rdb1608, cskrings, SB03}. The
multi-color images allow us to correct the positions and surface
brightness distributions of the lens galaxies for extinction, removing
a major source of systematic error common to previous modeling
efforts. Finally, we have measured the stellar velocity dispersion of
the dominant lens galaxy from a spectrum obtained with the Echelle
Spectrograph and Imager (ESI; Sheinis et al.\ 2002) on the Keck--II
Telescope. With this additional information and the improved
constraints on the time-delays, we are well-suited to refine the lens
models of this system and remove many of the major degeneracies and
systematic biases in the determination of~H$_0$.
 
This paper is organized as follows. In \S2 and \S3 we discuss the
observations and data reduction of the HST and Keck data,
respectively. In \S4, we describe the observational constraints
obtained from VLA and VLBA radio observations\footnote{The National
Radio Astronomy Observatory is a facility of the National Science
Foundation operated under cooperative agreement by Associated
Universities, Inc.}, HST images, and the Keck spectrum.  In \S5, we
briefly discuss a new lensing code and the $\chi^2$ minimization
procedure. In \S6 and \S7, the modeling results and inferred
value of H$_0$ are discussed. In \S8, we summarize and discuss our
results. Throughout this paper, we assume $\Omega_{\rm m}=0.3$ and
$\Omega_{\Lambda}=0.7$.

\section{Imaging}

The main objectives of the analysis of the multi-color HST imaging are
to obtain: (i) accurate extinction-corrected images of the two main
lens galaxies, from which their true centroids and structural
parameters (i.e., position angle, ellipticity and effective radius)
can be determined; and (ii) a deconvolved Einstein Ring after subtraction
of a model of the lens galaxies. The shape of the Einstein Ring places
additional constraints on the azimuthal structure of the lens model
\citep{cskrings}. We note that the analysis of the HST images
presented here is independent of that in Surpi \& Blandford (2003;
hereafter SB03), but produces nearly identical color maps.  Finally,
we bring all astrometric data obtained from the HST images to a
common ``reference frame'', i.e., that of the high-resolution VLBA
observations. This allows us to combine the radio and optical/infrared
observations in an self-consistent way, taking proper care of
potential differences (i.e., offsets, rotations and scalings) between
the different data sets.

\subsection{Observations and data reduction}

The HST images of the gravitational lens system B1608+656 are
available from the HST archive. Since the primary concern here is
obtaining high spatial resolution and an accurate model of the
point-spread function (PSF), we select the images with the best
resolution and sampling in each of the available bands (see also the
discussion in SB03): $41,344$s in 7 exposures with the Near Infrared
Camera and Multi Object Spectrograph (NICMOS) Camera 1 (NIC1) through
filter F160W (GO-7422; PI: Readhead); and 4 exposures on the Planetary
Camera (PC) of the Wide Field and Planetary Camera 2 (WFPC2) through
two filters for total exposure times of 11,500s and 11,600s in the
F814W and F606W bands, respectively (HST GO-6555; PI: Schechter).

The images are first reduced using a series of {\sc
iraf}\footnote{IRAF (Image Reduction and Analysis Facility) is
distributed by the National Optical Astronomy Observatories, which are
operated by the Association of Universities for Research in Astronomy
under cooperative agreement with the National Science Foundation.}
scripts based on the {\sc drizzle} package \citep{FH02}.  The scripts
clean and combine the images, by iteratively refining the measurement
of relative offsets and the identification of cosmic-rays and defects
\citep[similar to the scripts used in][Treu et al.\ 2003]{KT02}. The
final F160W ``drizzled'' image is shown in the top left panel of
Fig.1. Images of the system in the other filters are similar to the
ones shown in SB03 and are not repeated here.

\subsection{Dust extinction correction}

The B1608+656 system is characterized by two lens galaxies, G1 and G2,
with a strong color gradient in the region connecting the two
galaxies. The positions of the centroids of the two galaxies with
respect to the multiple images were observed to vary by as much as 86
milli-arcseconds (mas) between the F160W and F606W images
\citep[SB03;][]{KF99,F02}.

To remove this source of systematic uncertainty, we assume that the
observed color gradients arise from dust extinction (see SB03) and
that the two lens galaxies have the same intrinsic colors. We use the
available colors to measure the color excesses and correct for dust
extinction based on an extinction law (see below for a discussion). To
check for systematics, we perform the correction for each pair of
filters, in the reference frame of each one of the filters. The
results from each pair of filters are very similar and affect the
results only at a very small level, which will be taken into account
in the final error analysis on H$_0$. In the interest of space we will
describe as an example only the procedure for the F814W--F160W pair in
the F160W reference frame, mentioning the differences in the results
for the other frames where relevant. First, we deconvolve the F814W
image using a Lucy--Richardson algorithm and a PSF generated using
{\sc tiny tim} \citep{tinytim}. Second, we align the deconvolved F814W
image to the F160W image assuming that the positions of the point
sources are the same in each filter and not affected by
dust\footnote{Since the centroids of the lensed images are very sharp,
they change very little due to extinction (see also \S2.4).}, and that
the transformation is a rotation and translation with arbitrary axis
scaling. Third, the deconvolved and transformed F814W image is
convolved with a NICMOS PSF generated by {\sc tiny tim}, to match the
resolution of the F160W image. Fourth, the ratio of the matched images
in the two filters is used to produce a color map, which is found to
be very similar to the one shown in SB03. Fifth, it is assumed that
the intrinsic colors of G1 and G2 are spatially uniform and equal to
the minimum of the color measured in the region including G1 and G2
\citep[see also][]{SB03}.  Finally, the color excess is converted into
extinction in the individual filters using the following relations
computed using the Galactic extinction law by Cardelli et al.\ (1989)
for R$_{\rm V}=3.1$: A(F160W)/A$_{\rm V}$=0.415, A(F814W)/A$_{\rm
V}$=1.091, A(F606W)/A$_{\rm V}$=1.561. The extinction-corrected
image\footnote{Note that this correction is only made to the lens
galaxies and not to the Einstein Ring.} in F160W is shown in the top
right panel of Fig.~1.

This procedure is repeated for each pair of filters, in the reference
frame of each filter. We emphasize that this procedure does not assume
a particular smooth surface brightness profile for the lens galaxies,
nor does it force the centroids of the lenses to coincide in the
different filters. However, the production of a smooth
extinction-corrected surface brightness distribution for the lens
galaxies strongly suggests, although does not guarantee, that the
extinction correction has been done properly. Similarly, the
consistency of the astrometry between various filters provides an
additional test of the uncertainties related to the assumption of a
specific dust extinction law, the assumption of coincident centroids
of the lensed images, and the PSF modeling used for convolution and
deconvolution purposes. These tests are discussed in the next two
sub-sections, together with surface photometry and astrometry
measurements.

\subsection{Surface photometry}

Surface photometry is performed on the extinction-corrected F814W and
F160W images, as described in \citet{T99,T01b}. The F606W image is not
used for this task, given that it has the largest dust extinction and
the lowest signal-to-noise ratio (SNR), especially on G2.  The
galaxies are modeled as two \dv\ profiles -- which has been
shown to be a good description of the extinction-corrected surface
brightness profile of G1 \citep{rdb1608} -- that are fit
simultaneously to obtain effective radii ($R_e$) and total magnitudes
(Table~1). Only the region well inside the Einstein Ring is used in
the fit, to minimize contamination by the extended ring structure.
Very similar photometric parameters are obtained in the two filters,
demonstrating that uncertainties in the alignment of the images and in
the PSF modeling do not affect significantly the derived surface
photometry.  The best fitting model in filter F160W is shown in the
lower left panel of Fig.~\ref{fig:HST}.

To inspect the residuals from the fit, we apply the dust extinction
map to the best fitting \dv\ model and subtract the reddened model
from the original image. The brightness contribution of the galaxies
at the ring radius is found to be negligible. The residual image is
shown in the lower right panel of Fig.\ref{fig:HST}, demonstrating
that indeed the data are consistent with two relatively smooth
spheroids, reddened by a dust lane \citep[][]{SB03} without strong
evidence that the primary lens galaxy G1 is much affected by tidal
interactions with G2. The ring structure is very clear in the residual
image \citep[e.g.,][]{cskrings,SB03}, and is even more prominent in
the deconvolved residual image (Fig.5).

\bigskip
\begin{inlinetable}
\centering
\begin{tabular}{lcc}
\hline
\hline
                           & G1             & G2 \\
\hline 
 $F160W$ (mag)             & 16.41$\pm0$.08 & 18.18$\pm$0.05 \\ 
 $F814W$ (mag)             & 18.23$\pm$0.12 & 19.99$\pm$0.10 \\
 $R_{e,F160W}$ (arcsec)    & 0.62$\pm$0.05  & 0.27$\pm$0.03 \\  
 $R_{e,F814W}$ (arcsec)    & 0.58$\pm$0.06  & 0.26$\pm$0.04 \\
 $b/a$=$(1-e)$             & 0.5$\pm$0.1    & $\equiv1$\\
 Major axis P.A.($^\circ$) & $79\pm2$       & N/A \\
\hline
\hline
\end{tabular}
\end{inlinetable}

\noindent{Table~1 --- \small Observed photometric quantities of the
lens galaxies G1 and G2. All magnitudes are in the Vega system. The
systematic uncertainty on the NICMOS zero point \citep[see the NICMOS
instrument handbook;][]{2002hstd.book.....M} is not
included. Magnitudes are the total magnitudes obtained from fitting
the galaxy surface-brightness profiles with an $R^{1/4}$ law.
Magnitudes have been corrected for internal extinction, as discussed
in the text, but are not corrected for Galactic extinction. The axial ratio
of G2 is set to unity during the fit.
\label{tab:HST}}\medskip

\subsection{Astrometry}

The extinction-corrected images are used to measure the locations (in
pixels) of the centroids of G1 and G2, using the task {\sc imcentroid}
in {\sc iraf}. The deconvolved residual images are used to measure the
centroids of the multiple images of the lensed source, as well as the
shape of the ring (see \S4.2 and right panels in Fig.5).

Since we perform all measurements from the HST images in their
original pixel frames, we need to align the optical and infrared
frames to the much more accurate VLBA radio frame. We determine the
transformation (i.e., rotation, translation and scaling) that
minimizes the positional differences between the lensed image
centroids measured from the HST images, and their radio counterparts,
weighting properly by the measurement errors. We allow for different
pixel scales in the $x$/$y$ directions on the NIC1 and PC chips. We
find pixel scales of 0\farcs0457/0\farcs0456 for the PC and
0\farcs0434/0\farcs0431 for NIC1, in excellent agreement with the
values in the HST instrument handbook \citep{2002hstd.book.....M}.

The positional differences between the optical/infrared and radio
image positions are $\la 10$~mas in RA and DEC for each of the
individual lensed images. Since the transformation is determined from
the full set of four images, we estimate an error of $\sim$4~mas in
$x$ and $y$ on the transformation between the radio and NICMOS
frames. Using this transformation, we map the trace of the brightness
peak of the Einstein Ring to the radio frame (see \S4.2). Since the
transformation errors are negligible compared with the positional
determination of the brightness peak along spokes, one can assume for
all practical purposes that the radio data and the Einstein Rings are
perfectly aligned.

We use the same transformation to bring the galaxy centroids in
F160W to the radio frame. The combined centroid errors and
transformation errors quadratically add to an error of 6~mas in $x$
and $y$. We adopt this value as the centroiding errors of the galaxies
with respect to the lensed radio images. The measurement errors on the
radio positions are negligible \citep[$\ll$1~mas;][]{KF99}. We repeat the same
procedure for the F814W image.

After transformation to the radio frame, the positions of the lens
galaxies G1 and G2 agree to within 14 mas and 20 mas between the F160W
and F814W filters, respectively, as opposed to 72 mas and 38 mas in
the uncorrected frame. Slightly larger differences are found using the
F606W filter, as expected because this is the filter where
uncertainties in the dust corrections are the most severe\footnote{The
maximum difference in the centroid coordinates between F606W and
F160W is now 21 mas, with an rms scatter of 14 mas, as opposed to 86
and 45 mas respectively in the uncorrected images of SB03.}.  All
astrometric measurements from the F160W and F814W filters are listed
in Table~2. The deconvolved Einstein Rings in the F160W and F814W
filters are shown in the left panels of Fig.5.

\setcounter{table}{1}
\bigskip
\begin{table*}
\caption{Astrometry of the B1608+656 system}
\centering
\begin{tabular*}{15cm}{c@{\extracolsep{\fill}}rrrrrr}
\hline
\hline
  & \multicolumn{2}{c}{VLBA/8.4--GHz} & \multicolumn{2}{c}{HST/F814W} & \multicolumn{2}{c}{HST/F160W} \\\smallskip
 Image & $\Delta x$ ($''$) & $\Delta y$ ($''$) & $\Delta x$ ($''$) & $\Delta y$ ($''$) & $\Delta x$ ($''$) & $\Delta y$ ($''$) \\ 
\hline
 A &   $\equiv$0.0000  &  $\equiv$0.0000 &   $-$0.0003  &  $+$0.0009 & $-$0.0029  &  $+$0.0042   \\
 B &   $-$0.7380  &  $-$1.9612 &             $-$0.7376  &  $-$1.9588 & $-$0.7342  &  $-$1.9498   \\
 C &   $-$0.7446  &  $-$0.4537 &             $-$0.7455  &  $-$0.4528 & $-$0.7469  &  $-$0.4446   \\
 D &   $+$1.1284  &  $-$1.2565 &             $+$1.1298  &  $-$1.2563 & $+$1.1365  &  $-$1.2546   \\
\hline
 G1 &   N/A  &  N/A &  $+$0.4124 & $-$1.0592 & $+$0.4261 & $-$1.0581 \\
 G2 &   N/A  &  N/A &  $-$0.3091 & $-$0.9306 & $-$0.2897 & $-$0.9243 \\
\hline
\hline
\end{tabular*}
\end{table*}

\section{Spectroscopy}\label{sec:spec}

In this section we describe the spectroscopic observations of
B1608+656 with ESI. The main objective of the observations was to
obtain the stellar velocity dispersion of the primary lens galaxy
(G1), the measurement of which allows us to determine the power-law
slope of its radial mass profile and break the degeneracy with the
inferred value of the Hubble Constant.  Special care is required in
order to determine an unbiased stellar velocity dispersion and a
correct uncertainty, since G1 is a post-starburst galaxy.  This is
accomplished through extensive Monte Carlo simulations of artificial
spectra with different fractions of ``old'' and ``young'' stellar
populations.

\bigskip
\begin{inlinefigure}
\begin{center}
\resizebox{\textwidth}{!}{\includegraphics{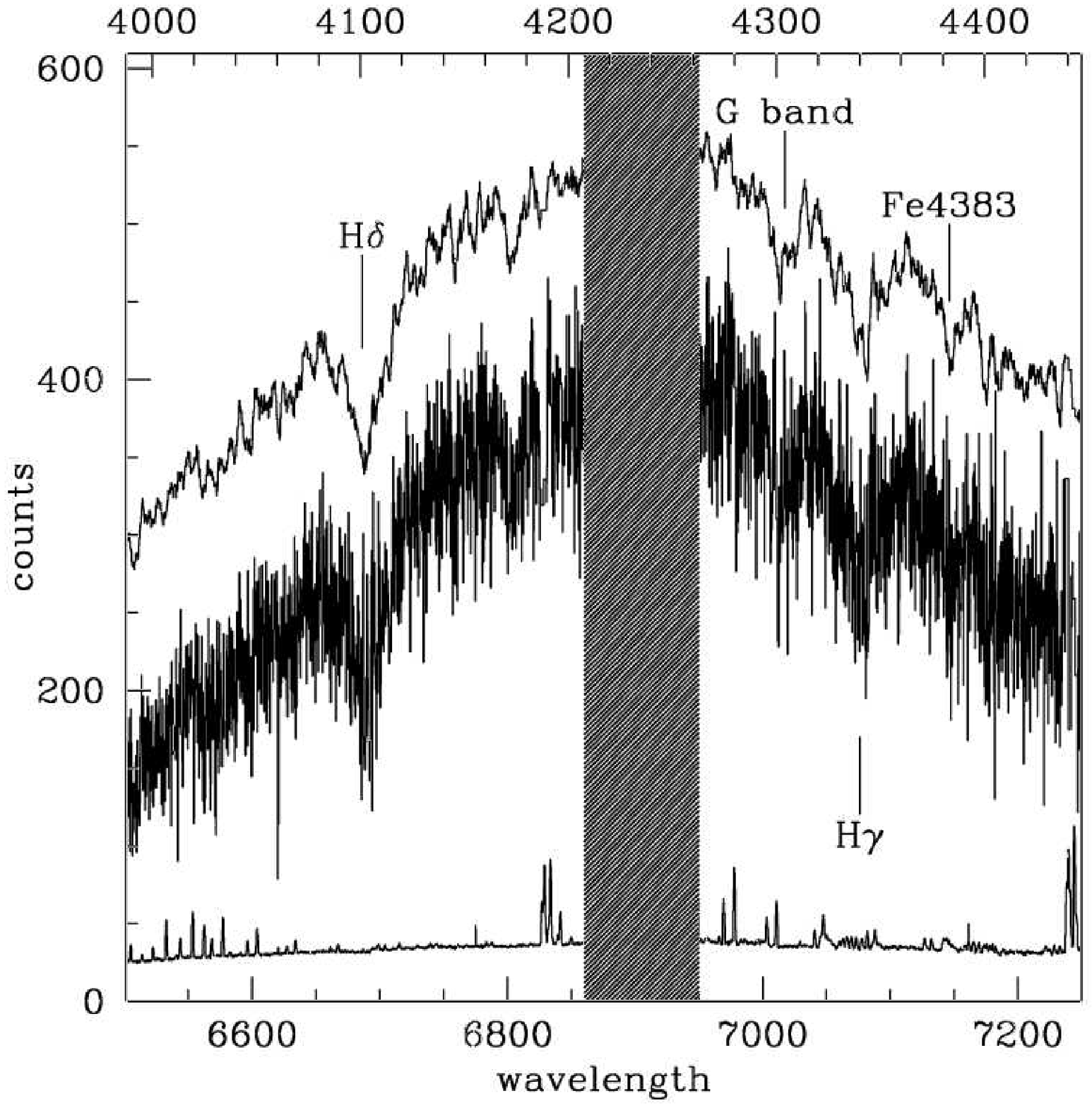}}
\end{center}
\figcaption{\label{fig:spec} Spectrum of B1608+656--G1 in one of the
echelle orders around the G--band (in counts, not flux
calibrated). The lower histogram shows the noise per pixel. A boxcar
smoothed version of the spectrum is shown above the original spectrum
to guide the eye. The wavelength range used for the kinematic fit was
6750--7200\AA\ (observed), which includes the G-band, H$\gamma$, and 
Fe4384 features, but excludes H$\delta$.  The atmospheric B-band 
(hatched band) and H$\gamma$ were masked out in the kinematic fit.}
\end{inlinefigure}

\subsection{Observations and data reduction}

We observed B1608+656 using ESI (Sheinis et al.\ 2002) on the Keck--II
Telescope on 2000 July 3, for a total integration time of 5400s
(3x1800s). The seeing was good with FWHM$\approx0\farcs85$ and the
night was clear. Between each exposure, we dithered along the slit to
allow for a better removal of sky residuals at the red end of the
spectrum. The slit (20$''$ in length) was positioned at PA$=83^\circ$,
i.e., aligned to within 4 degrees of the major axis of G1. The
slit-width of $0\farcs75$ yields an instrumental resolution of
$\sigma$$\sim$20\,\kms\ which is adequate for measuring the stellar
velocity dispersion and removing narrow sky emission lines. The
centering of the galaxy in the slit was constantly monitored by means
of the ESI viewing camera -- the galaxy was bright enough to be
visible in exposures of a few seconds duration -- and we estimate the
centering perpendicular to the slit to be accurate to better than
$0\farcs1$.

Data reduction was performed using the {\sc iraf} package
EASI2D\footnote{developed by D.~Sand and T.~Treu; Sand et al. (2003),
in prep.} as discussed in \citet{KT02} and Sand et al.\ (2002,3). A
one dimensional spectrum was extracted (Fig.~\ref{fig:spec}) by
summing the signal within an aperture corresponding to
$1\farcs7\times0\farcs75$. The SNR of the spectrum in the vicinity of
the G-band ($\sim$4304\,\AA) is 20\,\AA$^{-1}$. Note the very
prominent Balmer absorption features H$\gamma$ and H$\delta$, typical
of a K+A post starburst population \citep[Dressler \& Gunn
1983;][]{sm1608,SB03}.

\subsection{The Stellar velocity dispersion of G1}

The luminosity weighted velocity dispersion $\sigma_{\rm ap}$ of G1
was measured with the Gauss--Hermite Pixel-Fitting Software
\citep{vdm94} on the spectral region covering the observed wavelength
$\sim 6750-7200$ (Fig.~\ref{fig:spec}) excluding H$\delta$ and
including the G-band, H$\gamma$, and Fe4383 (Trager et al.\ 1998). As
kinematic templates we used spectra of G--K giants observed at
twilight with a $0\farcs3$ slit width, appropriately smoothed to match
the instrumental resolution of the $0\farcs75$ slit. The regions
around 6870 \AA, affected by atmospheric absorption, and the H$\gamma$
feature were masked during the fit. The result is robust with respect
to changes in the spectral region used in the fit, the fit parameters
(e.g., the order of polynomial used to reproduce the continuum level),
and the adopted template. The best fit gives: $\sigma_{\rm
ap}=247\pm35$ \kms\ in an aperture of 1\farcs7$\times$0\farcs75
centered on G1.

To minimize the amount of corrections to the data, we will only use
this measurement of the luminosity weighted velocity dispersion (as
opposed to the central velocity dispersion, defined in a standard
aperture) in the kinematic analysis (\S4), properly taking the seeing
and aperture effects into account by applying those corrections to the
models. However, to facilitate comparison with earlier work and other
applications of this measurement, we also give the central velocity
dispersion\footnote{The correction from luminosity weighted velocity
dispersion to central velocity dispersion is computed based on the
range of observed velocity dispersion profiles of local E/S0 galaxies,
and taking into account seeing effects as described in \citet{T01b}.},
$\sigma=269\pm40$ \kms, within a standard circular aperture of radius
$R_{\rm e}/8$.

We note here that if one adopts the measured central velocity
dispersion and the structural parameters measured from the extinction
corrected images (Table~1), to determine the offset of G1 from the
local Fundamental Plane, one finds the galaxy to be significantly
brighter than quiescent early-type of the same mass and at similar
redshifts. This is consistent with what is typically found for K+A
galaxies (e.g., van Dokkum \& Stanford 2003). In contrast, this result
differs from the findings of Rusin et al.\ (2003a), who estimated
a larger velocity dispersion from the image separation and measured a
lower luminosity for G1, resulting in an offset from the local
Fundamental Plane consistent with quiescent evolution.

The rest of this section describes Monte Carlo simulations aimed at
ensuring that our measurement provides not only an unbiased value of
the stellar velocity dispersion of G1, but also a realistic estimate
of the uncertainty.  In fact, although the SNR and resolution are more
than adequate to measure the velocity dispersion of a uniformly old
stellar population (e.g., Treu et al.\ 2001), the spectrum of G1 is
more challenging, because of the contamination by young stars.  Our
Monte Carlo approach works as follows. First, we model the spectrum as
a combination of a G8III star, to represent the old population, and of
an A5V star, to represent the young population (Dressler \& Gunn 1983;
Franx 1993). The stellar spectra are smoothed to a velocity dispersion
of 250~\kms, taking into account the initial spectral resolution of
the stellar spectra\footnote{The G8III star is from the ESI data, the
A5 V is taken from the library of Jacoby, Hunter \& Christen 1984; see
Treu et al.\ 2001 for a discussion of the resolution of that
library.}, and then combined with variable weights. Second, for each
set of weights (hereafter ``old'' fraction and ``young'' fraction)
1,000 realizations are constructed, adding noise to reproduce the SNR
of our data. Third, we run the Gauss--Hermite Pixel-Fitting Software
\citep{vdm94} on each realization to simulate the stellar velocity
dispersion measurement.

\bigskip
\begin{inlinefigure}
\begin{center}
\resizebox{\textwidth}{!}{\includegraphics{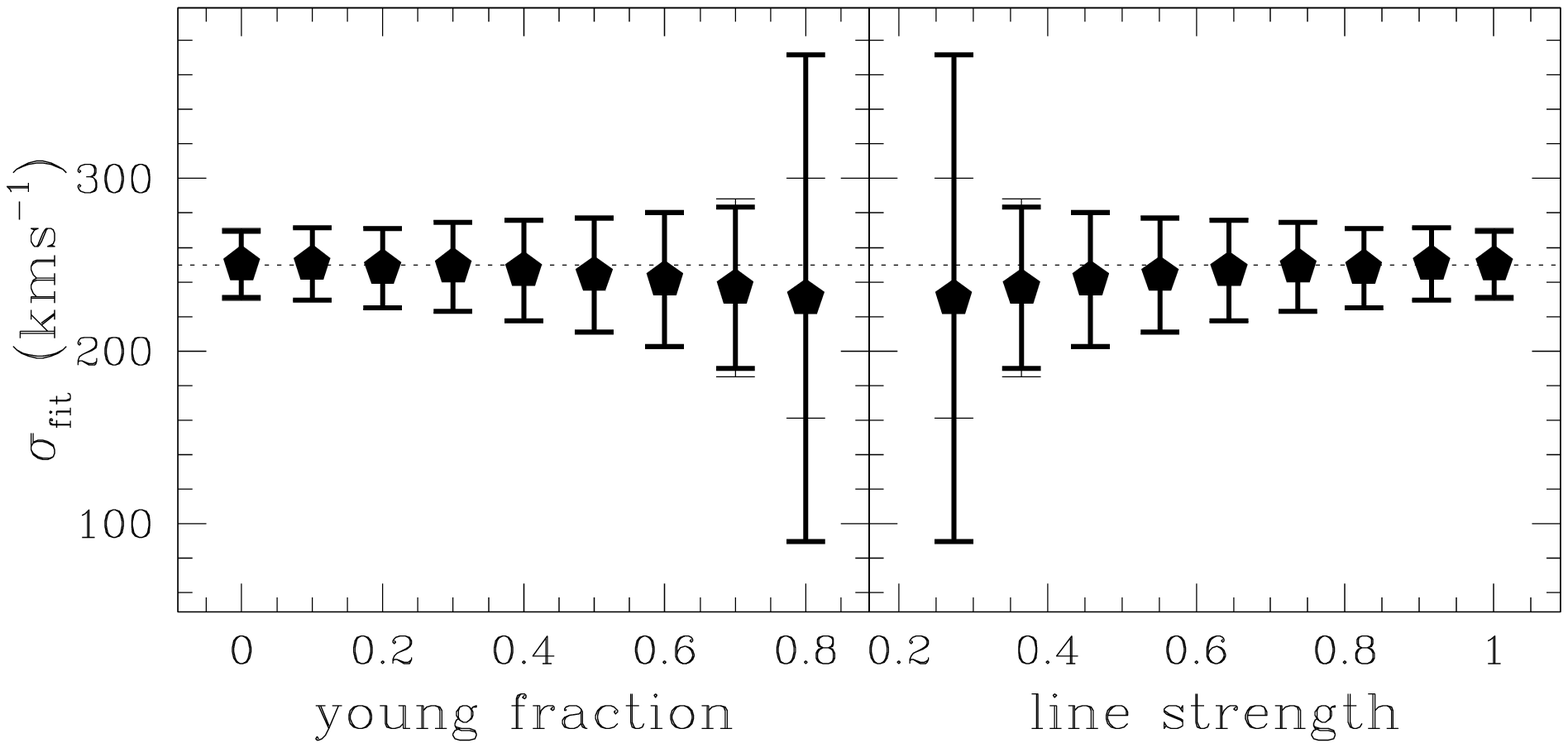}}
\end{center}
\figcaption{\label{fig:Monte} Monte Carlo simulations of the stellar
velocity dispersion recovery from artificial spectral of
B1608+656--G1. The dotted horizontal line shows the input value, the
pentagons shows the average value measured from 1,000
realizations. The heavy error bars show the average uncertainty
estimated by the code, while the light error bars show the
r.m.s. scatter of the recovered values}
\end{inlinefigure}

The results of the Monte Carlo simulations are shown in
Fig.~\ref{fig:Monte}. In the left panel we plot the average and
r.m.s. scatter (solid pentagons with thin error bars) of the recovered
velocity dispersion $\sigma_{\rm fit}$ as a function of the fraction
of A5 V light (young fraction). For young fractions up to $\sim0.6$,
the procedure recovers the input velocity dispersion to within
4\%. Only for a young fraction larger than 0.7 does the procedure
become biased. This happens because, as the young fraction increases,
the information content from the metal absorption features gets washed
away by the contribution of the young population.  Similarly, for
young fractions less than 0.6, the r.m.s. scatter (thin error bars)
agrees remarkably well with the average uncertainty returned by the
code (thick error bars). We can estimate the fraction of young stars
in the spectrum of G1 by comparing the observed depth of the G-band
and of H$\gamma$ with those in the model spectra. In
Fig.~\ref{fig:fit} we show that a composite spectrum with $\sim$40\%
of young population matches the observed spectrum, while fractions
$\sim$70\% and higher produce much deeper $H\gamma$ absorption lines
than observed. We therefore conclude that the fraction of young
stellar light in G1 must be smaller than $\sim$70\% and that the value of
$\sigma_{ap}$ is the correct and unbiased measurement of the stellar
velocity dispersion of G1. Similarly the error estimate is unbiased.

\bigskip
\begin{inlinefigure}
\begin{center}
\resizebox{\textwidth}{!}{\includegraphics{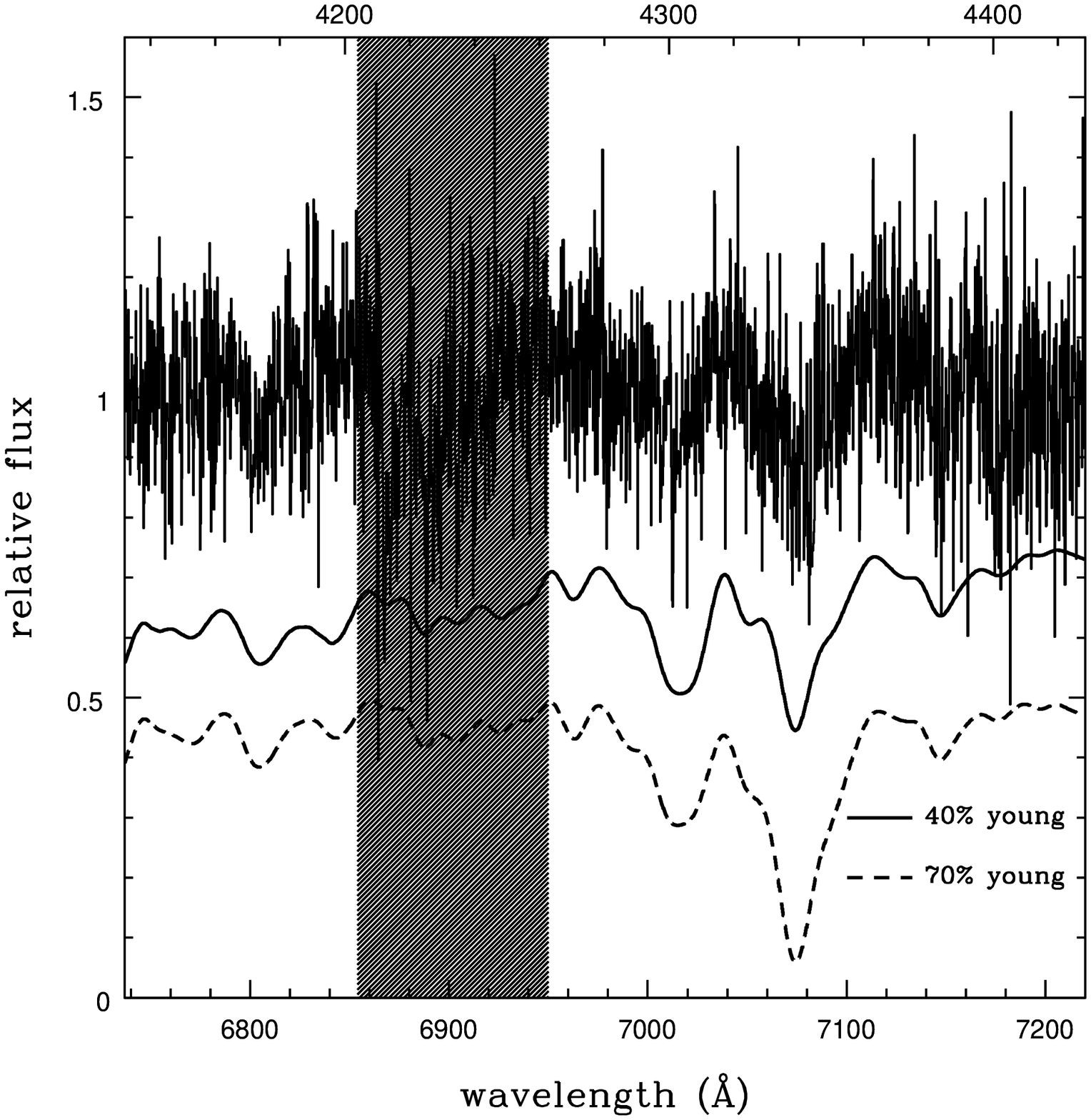}}
\end{center}
\figcaption{\label{fig:fit} A fit of $\sigma_{\rm ap}$ for G1 in
B1608+656 for two different fractions of the young stellar population
(shifted for clarity). A young fraction of 70\% clearly overpredicts
the depth of the H$\gamma$ feature.}
\end{inlinefigure}

A more quantitative confirmation of this result comes from the
inspection of the line strength, which is used by the Gauss--Hermite
Pixel-Fitting Software \citep[see definition in][]{vdm94} to measure
the depth of the absorption features: the larger the line strength,
the deeper the absorption features.  In the models, by construction,
the line strength of the old stellar population anticorrelates very
well with the young fraction. In fact a very similar conclusion is
drawn if line strength is used as independent variable to show the
result of the Monte Carlo simulations (right panel of
Fig.~\ref{fig:Monte}). When the line strength of the old stellar
population is larger than 0.4 (corresponding approximately to a young
fraction smaller than 0.7) the signal to noise ratio and resolution of
our spectrum are sufficient to obtain an unbiased measurement of the
stellar velocity dispersion.  For a real composite population the
interpretation of this parameter is more complex, since it depends on
the chemical abundances and possible additional contributions to the
continuum. Nevertheless, it provides a useful quantitative measure of
the depth of the metal absorption features. which is the main
parameter that controls the accuracy of the fit. The best fitting
template yields line strength $0.5\pm0.05$, also confirming that the
measurement of velocity dispersion is unbiased.

\section{Mass Model and Observational Constraints}

We model the lensing and kinematic properties of both lens galaxies,
G1 and G2, assuming that their luminous plus dark-matter mass
distributions are elliptical (i.e., oblate in three
dimensions\footnote{Some early-type galaxies show evidence for
iso-surface-density twists that could be due to triaxial mass
distributions.}) and have a radial power-law dependence, (i.e.,
$\rho\propto r^{-\gamma'}$). In particular, we assume for the lens
models a dimensionless surface mass density of
\begin{equation}
   \kappa(x,y) = {b}{\left[x^2 + (y/q_{\ell})^2 \right]^{(1-\gamma')/2}},
\end{equation}
where $q_{\ell}$ is the axial ratio of the surface mass distribution
and $\gamma'$ the slope of the corresponding density distribution
($\gamma'=2$ is designated as ``isothermal'' henceforth).  The mass
distribution has its centroid at $(x_\ell, y_\ell)$ and a major-axis
position angle of $\theta_{\ell}$ (measured north through east).
In addition, we allow for a single external shear for the system with
strength $\gamma_{\rm ext}$ and position angle $\theta_{\rm ext}$.

We choose to use power-law mass models, since they have been
successful in reproducing many of the detailed lensing observations
\citep[e.g.,][]{1995ApJ...445..559K, 2001ApJ...554.1216C,
2001ApJ...558..657M, 2002MNRAS.330..205R, 2003ApJ...587...80W} and
more recently also kinematic observations \citep[e.g.,][]{TK02a,KT03}
of lens systems and lens ensembles
\citep[e.g.,][]{2003ApJ...587..143R}, even though we acknowledge that
this choice might not be unique.  We feel at this point, however,
that there is no strong incentive to go beyond these models. From a
practical standpoint, including the constraints from the Einstein Ring
becomes computationally very expensive for more complex lens models
(see \S5 for a discussion).

For the dynamical modeling (i.e., determining the stellar velocity
dispersion), we solve the spherical Jeans equations, assuming a
spherical mass distribution with the same power-law index $\gamma'$ as
used for the lens models (i.e., Eq.(1)). We model the luminous mass
distribution as a trace population -- embedded in the luminous plus
dark-matter potential -- with a Hernquist luminosity-density profile
\citep{1990ApJ...356..359H}. We have also examined
\citet{1983MNRAS.202..995J} models and find differences typically
$\la$1\% for the models that we explore. Henceforth, we only use
the Hernquist profile, which closely follows a \dv\ brightness profile
over the radial range of interest and is analytically tractable. Since
the break radius of the dark-matter halo is well beyond the Einstein
radius, we assume it to be much larger than the latter, with
negligible effects on the calculation of kinematic quantities
\citep[see][for a full discussion]{TK02a,KT03}. Finally, we allow for
anisotropy of the stellar velocity ellipsoid, modeling the radial
anisotropy parameter $\beta$ \citep{1987gady.book.....B} with a
Osipkov--Merritt \citep{1979PAZh....5...77O,M85a,M85b} model
with anisotropy radius $r_i$. For $r_i=\infty$, the stellar velocity
dispersion is fully isotropic. We also take the seeing and aperture
into account. We refer to \citet{KT03} for a further discussion of our
kinematic model.

We now continue with a more detailed description of the precise set of
constraints that we use in the lensing and kinematic models that
are presented in \S6.

\subsection{The Radio and Optical Image Constraints}

The positions of the four lensed radio images can most accurately been
determined from high-resolution radio observations. We use the VLBA
positions listed in \citet{KF99}, which are consistent with VLA
observations to within $\sim$1~mas \citep{F02}. We choose positional
errors on the VLBA image positions of 1~mas, which are considerably
larger than the formal measurement errors (tens of
micro-arcseconds). This error is approximately that by which the most
massive mass substructures in the lens (expected to be on the order of
$\sim 10^9$~M$_\odot$) could potentially change the image
positions. Since mass substructures are ``randomly'' distributed such
positional shifts can practically be regarded as random errors.

Radio flux ratios might not be reliable constraints on macro lens
models; either because of mass substructure (e.g. Mao \& Schneider
1998; Metcalf \& Madau 2001; Keeton 2001; Chiba 2002; Metcalf \& Zhao
2002; Dalal \& Kochanek 2002; Brada\v{c} et al. 2002; Keeton et
al. 2003; Brada\v{c} et al. 2003), radio microlensing
\citep[e.g.,][]{2000A&A...358..793K, 2002ApJ...580..685S} or
inter-stellar-medium (ISM) propagation effects (Koopmans et al. 2003).
Hence, we conservatively adopt a large error of 20\% on the individual
radio fluxes (i.e., $\sim$28\% on the flux ratios). This is supported
by the fact that the formal errors on the flux ratios within a single
season are much smaller than the differences between seasons
\citep{F02}.  For the flux ratios we adopt the approximate median
value from three independent monitoring seasons of B1608+656
\citep{F02}. Since previous models had difficulties reproducing the
flux ratio of image D \citep{KF99,F02}, it is not used as a
constraint. Hence, if the refined lens models can recover the flux of
image D, we would have additional confidence in the lens models.

Finally, we use the time delays between the VLA images as constraints
\citep{F99,F02}. Time-delay ratios are particularly strong constraints
on the mass profile, since they are almost a direct measure of the
depth of the potential at the lensed-image positions. Since there are
three independent time delays and one Hubble Constant to solve for,
they add two constraints on the mass model (we emphasize here that
time delays are affected by substructure only at a negligible level,
$\ll$1\%). All point-image constraints are listed in Table~3.

\begin{table*}[t]
\caption{Galaxy centroid positions and point-image constraints on the B1608+656 system}
\centering
\begin{tabular}{crrrr}
\hline
\hline
 Image\tablenotemark{a,b,c} & $\Delta x$ ($''$) & $\Delta y$ ($''$) & $S_{\rm norm}$ & $\Delta t$ (days) \\ 
\hline
 A &   $\equiv$0.0000$\pm$0.001  &  $\equiv$0.0000$\pm$0.001 & 2.020$\pm$0.404 & 31.5$\pm$1.5 \\
 B &   $-$0.7380$\pm$0.001  &  $-$1.9612$\pm$0.001 &  1.000$\pm$0.200 & $\equiv$0.0 \\   
 C &   $-$0.7446$\pm$0.001  &  $-$0.4537$\pm$0.001 &  1.034$\pm$0.207 & 36.0$\pm$1.5 \\          
 D &   $+$1.1284$\pm$0.001  &  $-$1.2565$\pm$0.001 &  0.347$\pm$$\infty$ & 77.0$\pm$1.5 \\          
\hline
 G1 &  $+$0.4261$\pm$0.006 & $-$1.0581$\pm$0.006 & N/A & N/A \\
 G2 &  $-$0.2897$\pm$0.006 & $-$0.9243$\pm$0.006 & N/A & N/A \\
\hline
\hline
\end{tabular}
\tablenotetext{a}{The positions for components A-D are from
\citet{KF99}.}  
\tablenotetext{b}{The time delays and flux ratios are from \citet{F02}.}
\tablenotetext{c}{The G1 and G2 positions are from this paper.}
\end{table*}

\subsection{The Einstein--Ring Constraints}

The F160W image of B1608+656 exhibits a fully-connected Einstein Ring
around the two lens galaxies \citep{cskrings,SB03}. The structure of
Einstein Rings can be used as additional constraint on the
gravitational lens models \citep{rdb1608,cskrings}.  In particular,
the unit vector along a radial spoke at the position of the brightness
peak (on the spoke), when projected on the source plane, is
perpendicular to the gradient of the surface brightness distribution
of the unlensed source \citep{cskrings}. When the isophotes of the
source are assumed to be elliptical -- often a reasonable assumption
-- the trace of brightness peaks (along radial spokes) places extra
constraints on the lens potential at the minimal cost of adding four
additional free parameters (ring-source position, ellipticity and
position angle). Moreover, for an elliptical source, the trace of the
ring is independent of assumptions on the radial behavior of the
surface brightness distribution of the source.

Even though the SNR decreases with increasing distance from the
central nucleus of the lensed source along the ring, it is still
sufficient to fully trace the ring (see Fig.1 and Kochanek et al.\
2001). This can be done even more accurately on the deconvolved image,
where the Einstein Ring is ``sharper'' in the radial direction
(Fig.5). We note that color gradients over the image have very little
effect on the brightness peaks of the sharp ring \citep[see
also][]{cskrings}. To trace the deconvolved Einstein Ring, we define
90 independent spokes, separated by 4 degrees in angle (consistent
with the improved image resolution after deconvolution), radiating
from a point roughly in the middle of the ring. The origin of the
spokes is arbitrary. The brightness distribution of the deconvolved
ring along each spoke is subsequently fitted with a Gaussian. From
this fit, the position of the brightness peak and the FWHM of the ring
are obtained. The resulting fitted ring is shown in Fig.5. To ensure
that the ring is not significantly affected by uncertainties produced
by the deconvolution, we repeat the same procedure for the F814W
image. We assume that the error on the ring peak is proportional to
the ring width, since it is hard to estimate an error from a
deconvolved image. We adopt 0.42$\times$$\sigma_{\rm r}$ as error for
F160W (with $\sigma_{\rm r}$=FWHM/2.35 from the Gaussian fit), which
gives $\chi^2$/DOF$\sim$1, only weakly dependent on the particular
lens model. This particular choice is not crucial, however, and
somewhat higher or lower scale factors can be chosen without affecting
the modeling results. Hence, we put more weight on those parts of the
ring that are sharpest (near the radio images) and less weight on the
less well-defined parts of the ring. Notice the remarkable agreement
between the traces of the ring in F160W and F814W (Fig.5), lending
credit to the trace \citep[compare also to][who find a very
similar trace from the original, not deconvolved, ring]{cskrings}.

\subsection{Constraints from the Lens Galaxies}

As for the Einstein Ring, we use the transformation discussed in \S2.4
to bring the positions of the two lens galaxies in the F160W and F814W
images to that of the VLBA frame with positional errors of 6~mas with
respect to the radio images (Table~3). The good agreement of the
galaxy centroids from different filters -- after extinction correction
-- seem to indicate that the extinction corrections applied to the
F160W image have been done properly. Confidence in the extinction
correction is further supported by the fact that, after correction, G1
is well described by a $R^{1/4}$ brightness profile (see also
Blandford et al.\ 2001).  It seems highly unlikely that an improper
extinction correction would produce a well-behaved smooth surface
brightness distribution from an originally patchy image. In the lens
models, we therefore adopt the light centroids from the
extinction-corrected F160W images, transformed to the radio frame, as
the final values for the mass centers of G1 and G2.

In addition, modeling of the extinction-corrected brightness
distribution gives a position angle of (79$\pm$2)$^\circ$ (north to
east) of G1. Since the extinction-corrected galaxies appear
relatively smooth and unperturbed it is safe to assume that G1 is most
likely not much affected by tidal interactions with G2. The position
angle of G1 is used as a prior on the position angle of its mass
distribution.

As a final model constraint, we use the stellar velocity dispersion of
G1, $\sigma_{\rm ap}=247\pm 35$~\kms\ (\S3,2), which provides
information on the slope of its radial mass profile.

\section{Lensing Code and $\chi^2$ minimization}

To model the B1608+656 system, we have developed a new lens code,
highly optimized for speed and accuracy. This is necessary because
most of the models require $\chi^2$--minimization in a 22-dimensional
parameter space (see \S 6.1 for a listing of all free model
parameters). Since the structure of the Einstein Ring is included as a
constraint, each iteration (i.e., $\chi^2$ calculation) is
computationally expensive, especially if there are no analytic
solutions for the lens potential. Furthermore, a full non-linear error
analysis of the parameters of the lens model requires a repetitive
minimization of the $\chi^2$ function for different sets of fixed lens
parameters.

To maximize speed, the code is first optimized in its use of memory,
which can be critical when copying large data structures (i.e., those
containing all constraints and intermediate results). Second, analytic
solutions are used wherever possible, as well as publicly-available
fast algorithms developed for specific power-law mass models
\citep[e.g.,][]{1998ApJ...502..531B, 1998ApJ...506...80C,
2002ApJ...568..500C}. A significant speed-up of the $\chi^2$
minimization and error analysis is gained by using the {\tt MINUIT}
minimization package from CERN (James 1998), which has been optimized
for finding function minima in large-dimensional parameter spaces with
a minimum number of function evaluations.  We have compared the
results from the new code, for B1608+656 and various simple test
systems, with those from a previously used code \citep[e.g.,][]{KF99}
and from the publicly-available code from \citet{lensmodel}. We find
that they all agree to within the expected numerical errors.

The increase in speed of the new code over the other codes, when we
include the Einstein Ring data for B1608+656, can be several orders of
magnitude, depending on the specific situation. For example,
$\chi^2$--minimization with 22 free parameters usually takes minutes,
compared to hours for the code by \citet{lensmodel}, when we use mass
models that are not analytically tractable (e.g., non-isothermal
power-law density profiles). A full non-linear error analysis for all
parameters -- a built-in option with {\tt MINUIT} -- takes several
hours with the new code, but only several minutes extra for a single
parameter (e.g., H$_0$). A similar analysis with the code by
\citet{lensmodel} is expected to take several days and is clearly not
feasible for systems of this complexity.

To find the global $\chi^2$ minimum, we first optimize all free model
parameters using only the radio data as constraints. This rapidly
leads to a model that fits all the radio constraints extremely
well. In the second step, the Einstein Ring data are added to the
constraints. Since the mass model parameters are already relatively
close to the best solution, it takes only several minutes to converge
to the $\chi^2$ minimum. To check that indeed the correct global
$\chi^2$ minimum has been found, the models are started from different
initial conditions and also using different minimization techniques
(i.e., simulated annealing and downhill simplex methods).  Monte Carlo
minimization is done once a $\chi^2$ minimum is reached. Finally, the
minimum in all parameter-space directions is traced while performing a
non-linear error analysis\footnote{The non-linear error analysis
consists of re-minimizing $\chi^2$ for a range of fixed values for the
parameter of interest \citep{james94}.  The 68\% confidence limits,
quoted in this paper, correspond to the parameter range with $\Delta
\chi^2=1$ w.r.t.  the minimum--$\chi^2$ solution.}. None of this has
led us to suspect that the global $\chi^2$ minimum has not been
found. This is also supported by the fact that the observed lens
properties are reproduced well by the lens models.

\bigskip
\begin{table*}[t]
\caption{Lens mass model parameters for B1608+656.}\label{tab:isomodel}
\centering
\begin{tabular}{p{2.5cm}rrrrrrrr}
\hline
\hline
     & SIE &  SPLE1 &  SPLE2  & SPLE1+D &   SPLE1+D & SPLE2+D & SPLE2+D \\
     &     &       &        &  (iso) &  (aniso) & (iso)  & (aniso)            \\
\hline 
$x_{\ell}$ (arcsec) & 0.425 & 0.426 & 0.425 & 0.425 & 0.425 & 0.425 & 0.426 \\
             & $-$0.291 &  $-$0.290 & $-$0.290 & $-$0.291 & $-$0.290 & $-$0.291 & $-$0.290\\
$y_{\ell}$ (arcsec) & $-$1.071 & $-$1.069 & $-$1.069 & $-$1.069 & $-$1.069 & $-$1.069 & $-$1.069 \\
             & $-$0.929 & $-$0.928 & $-$0.928 & $-$0.928 & $-$0.928 & $-$0.928 & $-$0.928 \\
$b$ (arcsec$^{\gamma'-1}$)& 0.531 & 0.553 & 0.553 & 0.526 & 0.555 & 0.531 & 0.553\\
             & 0.288 & 0.263 & 0.263 & 0.269 & 0.263 & 0.268 & 0.263 \\
$q_{\ell}$ & 0.606 & 0.607 & 0.606 & 0.604 & 0.603 & 0.605 & 0.606\\
    & 0.340 & 0.308 & 0.308 & 0.318 & 0.307 & 0.316 & 0.308 \\
$\theta_{\ell}$ ($^\circ$) & 76.8 & 77.0 & 77.0 & 77.0 & 77.0 & 77.0 & 77.0\\
                    & 69.2 & 68.5 & 68.5 & 68.4 & 68.4 & 68.4 & 68.5 \\
$\gamma'$ & $\equiv$2.00 & 1.99$^{+0.20}_{-0.20}$ & 1.99$^{+0.14}_{-0.15}$ & 2.05$^{+0.14}_{-0.13}$ & 2.00$^{+0.14}_{-0.15}$ & 2.04$^{+0.11}_{-0.12}$ & 2.00$^{+0.12}_{-0.12}$ \\
          & $\equiv$2.00 & 2.12 & 2.12 & 2.12 & 2.12 & 2.12 & 2.12 \\
\hline
$\gamma_{\rm ext}$ & 0.085 & 0.081 & 0.081 & 0.077 & 0.081 & 0.078 & 0.081\\
$\theta_{\rm ext}$ ($^\circ$) & 6.8 & 10.6 & 10.6 & 13.4 & 10.4 & 12.9 & 10.6 \\
\hline
$x_s$ (arcsec) & 0.095 & 0.103 & 0.102 & 0.088 & 0.102 & 0.090 & 0.102 \\
$y_s$ (arcsec) & $-$1.070 & $-$1.070 &  $-$1.070 & $-$1.069 & $-$1.070 & $-$1.069 & $-$1.070 \\
\hline
$q_h$ & 0.639 & 0.640 & 0.640 & 0.634 & 0.640 & 0.635 & 0.640 \\
$\theta_h$ ($^\circ$) & 113.2 & 113.0 & 113.0 & 112.1 & 113.1 & 112.3 & 113.0 \\ 
\hline 
H$_0$~(km/s/Mpc) & 71$^{+5}_{-6}$ & 74$^{+10}_{-11}$ & 74$^{+7}_{-6}$ & 76$^{+7}_{-6}$ & 74$^{+7}_{-6}$ & 75$^{+6}_{-5}$ & 74$^{+6}_{-5}$\\ $\chi^2$ & 99.8 & 98.0 & 98.0 & 98.1 & 98.0 & 98.1 & 98.0\\ 
\hline 
\hline 
\end{tabular}\\\smallskip
\tablenotetext{}{Note: The first (second) line for each parameter
denotes values for G1 (G2). The 68\% confidence limits on H$_0$ and
$\gamma'_{\rm G1}$ have been determined from a full non-linear error
analysis, where all free parameters are varied for a range of fixed
values of these parameters until $\chi^2$ increases by unity. We limit
ourself to these two most important parameters, since this calculation
is computationally very expensive.}
\end{table*}

\section{Results}

In this section, we describe the results from the lens models. After
listing the free parameters and adopted priors of the general model in
\S6.1, we consider in \S6.2 a simplified problem in which the slopes
of the galaxies are fixed to isothermal. This is done to facilitate
comparison with previous work and to illustrate the effect of
correcting the centroid positions for dust extinction. In Section 6.3
we explore the full range of parameter space, and hence its effects on
H$_0$, allowing the slopes to vary to reproduce the lensing
geometry. Finally, in Section 6.4 we add the dynamical constraints to
reduce the uncertainties and break the mass profile degeneracy of
G1. All results have been calculated using the new lens code and are
based {\sl only} on constraints from the radio data, the NICMOS F160W
images, and the stellar kinematics. Constraints from the WFPC2 F814W
images will be used as additional check on systematic effects in
Section 7.3.

\begin{figure*}[t]
\begin{center}
  \leavevmode
\vbox{%
\hbox{%
  \epsfxsize=8.5cm
  \epsffile{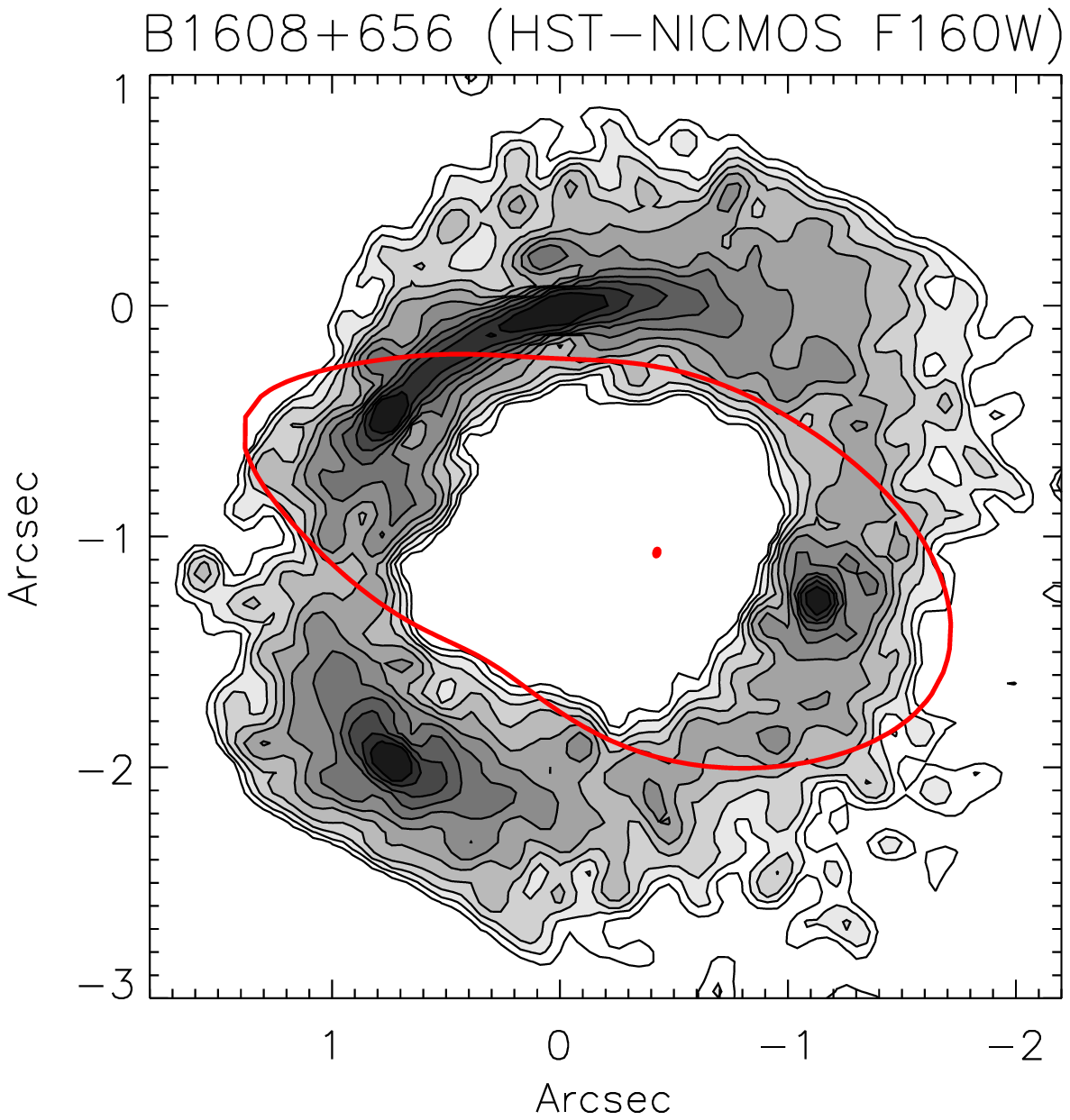}
  \epsfxsize=8.5cm
  \epsffile{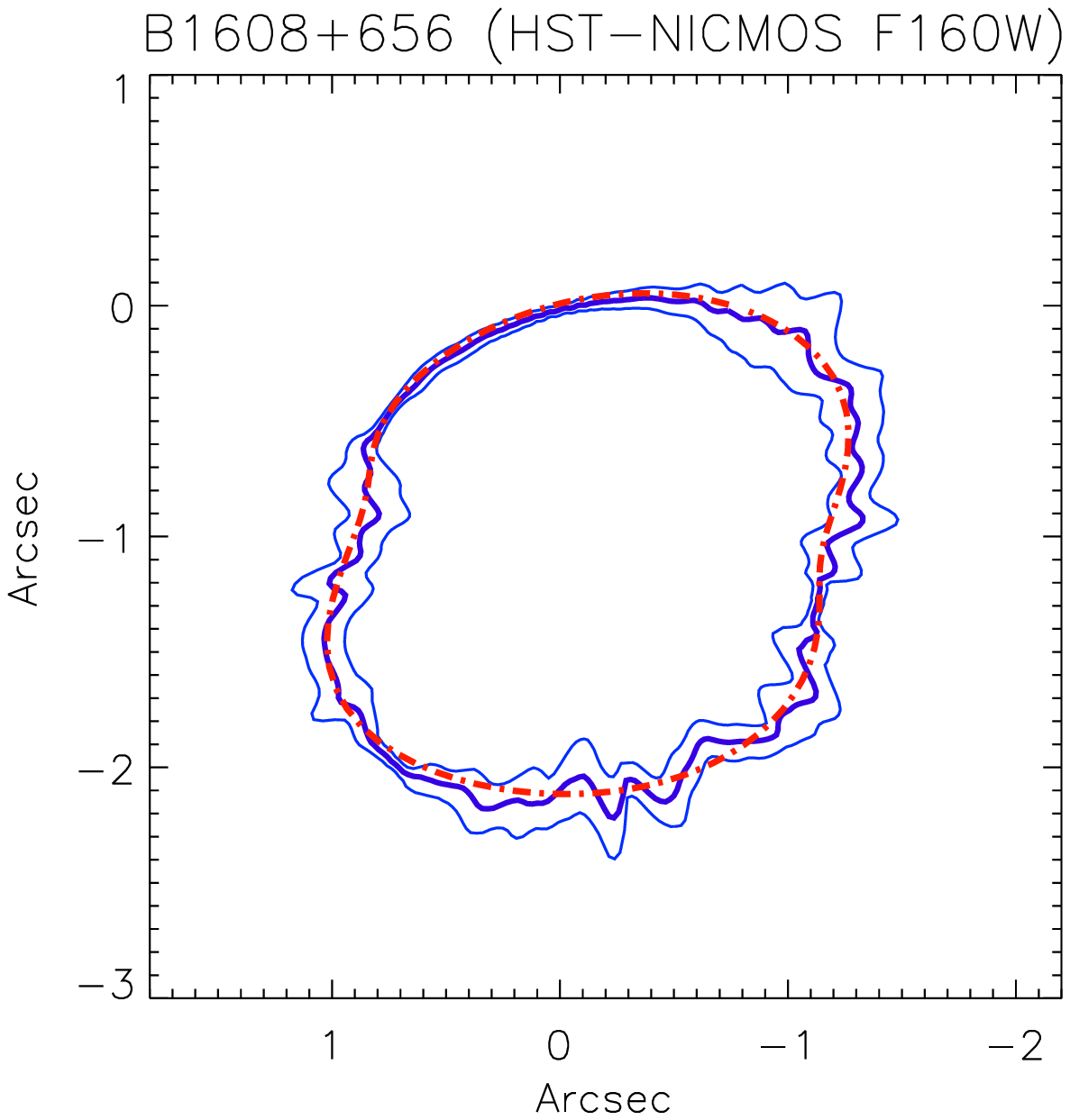}}
\hbox{%
  \epsfxsize=8.5cm
  \epsffile{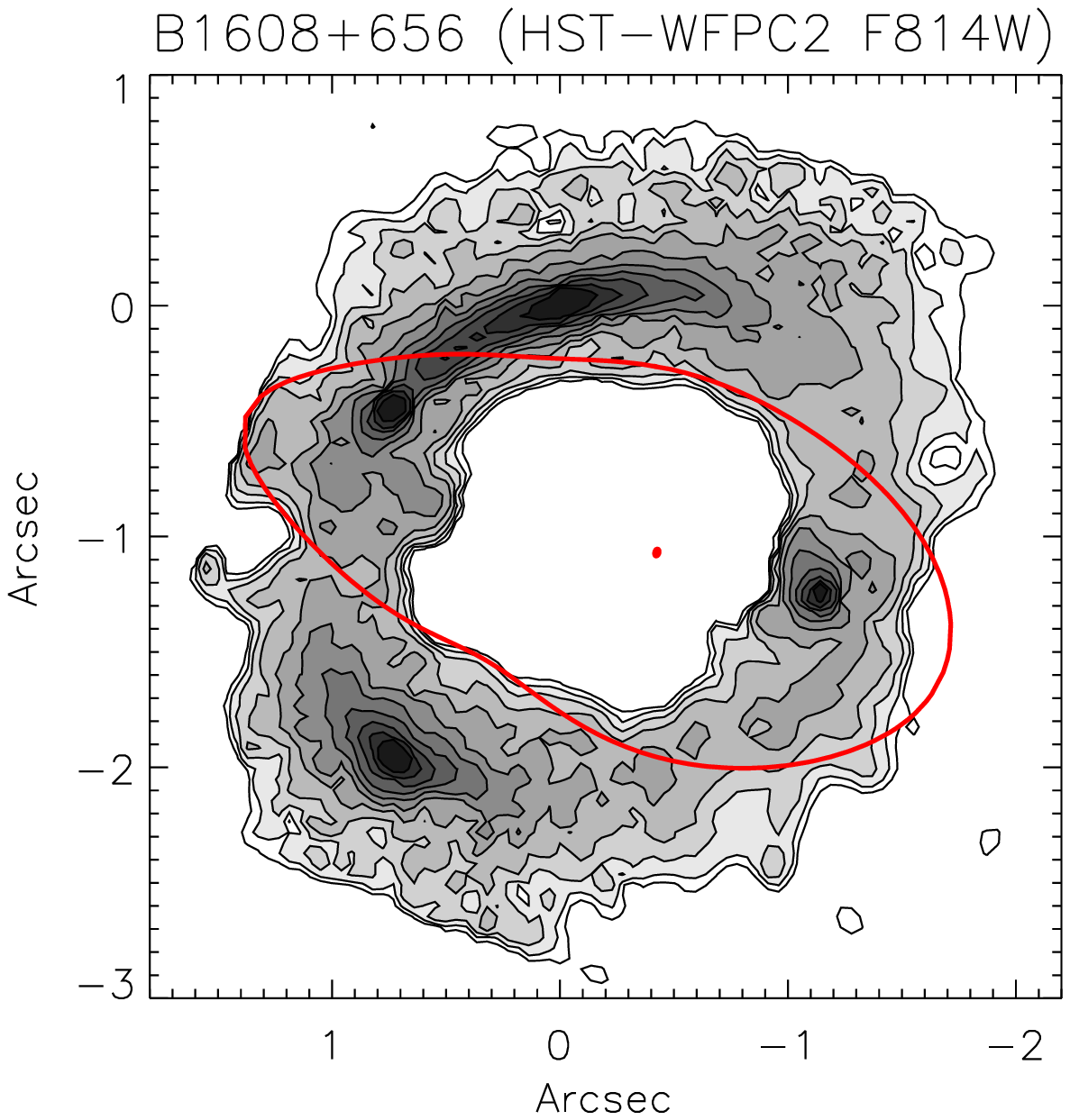}
  \epsfxsize=8.5cm
  \epsffile{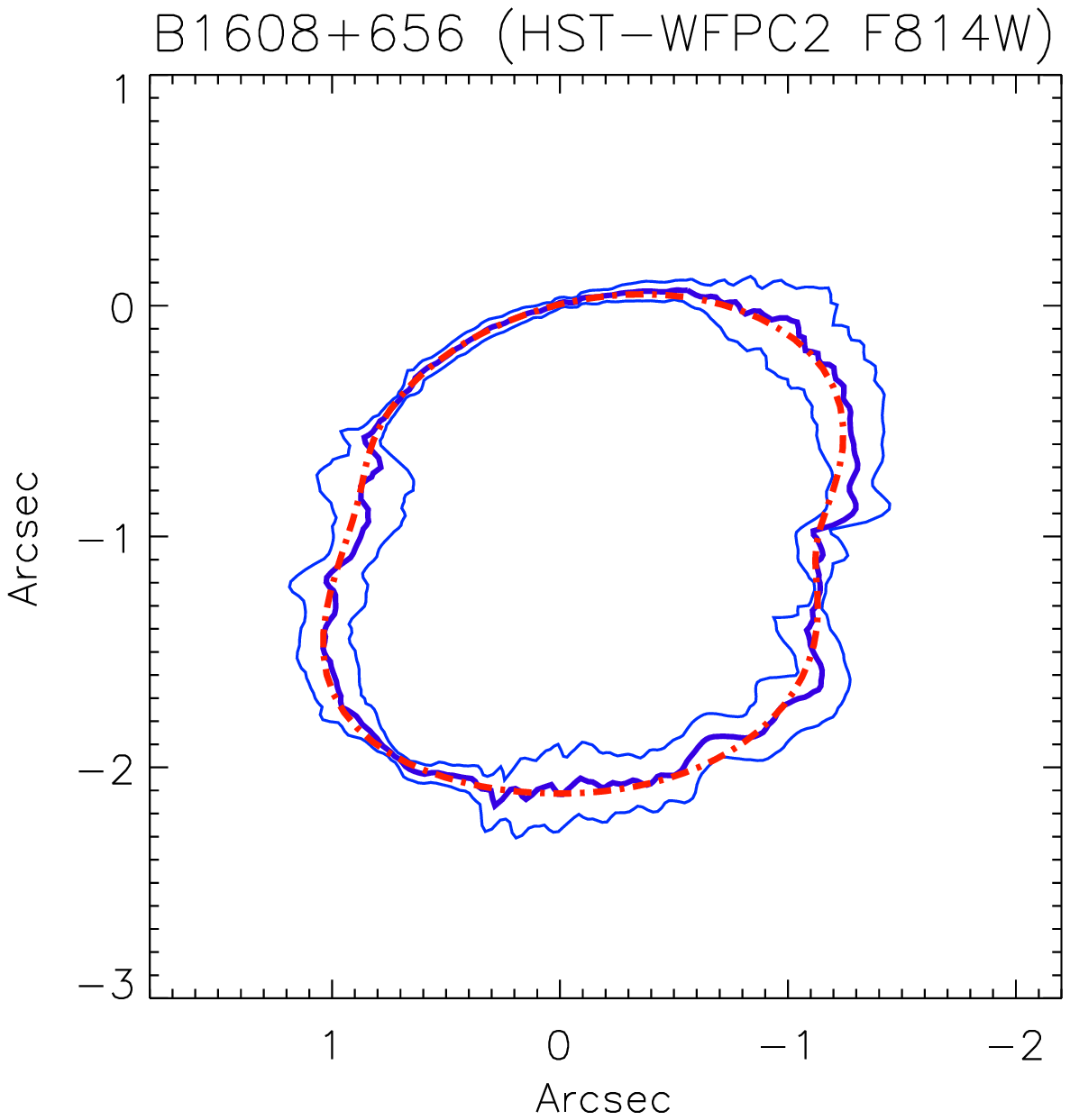}}}
\end{center}
\caption{The deconvolved Einstein Rings of B1608+656 in the F160W
(upper left) and F814W (lower left) bands, after extinction correction
and subtraction of a model of the two lens galaxies. The contour
levels increase by factors of two (arbitrary units). The solid curve
indicates the outer critical curve (see text) of our best lens model
(\S 6.4). The right panels indicate the trace of the Einstein Ring
(thick solid line) and the $\pm$1--$\sigma_{\rm r}$ width of the ring
(two thin solid lines). To smooth the trace, we plot it for 360 spokes
(only 90 are used in the modeling; see text). The dot-dashed curves
show the minimum--$\chi^2$ SPLE1+D models.}
\end{figure*}

\subsection{Free Parameters and Priors}

To allow maximum freedom in the lens models and a proper calculation
of the significance of the derived value of the Hubble Constant, we
allow nearly all of the model parameters (i.e., position, lens
strength, position angle, ellipticity, and density slope) of each lens
galaxy to vary, constraining several with Gaussian priors. The source
adds three additional free parameters (i.e., position and flux),
the Hubble Constant adds one, and the external shear adds two. The model of
the elliptical source that describes the Einstein Ring (\S4.2) adds an
additional four free parameters (position, ellipticity and position
angle). In total there are up to 22 free parameters.

We use Gaussian priors on the galaxy positions, placing them at their
observed positions and allowing for 1--$\sigma$ errors of 6~mas in $x$
and $y$ (Table~3). Hence, $\chi^2$ increases by unity if the galaxy
positions deviate by 1--$\sigma$ from the observed
positions. Similarly, we allow the position angle of G1 to vary using
a Gaussian prior of 79$\pm$2 degrees. Since B1608+656 is part of a
small group \cite[e.g.,][]{2002AAS...201.8009F} we allow for an
external shear. The group is not massive and has a low velocity
dispersion (Fassnacht et al.~in~prep.). We therefore place a
conservative 0.00$\pm$0.10 (1--$\sigma$ positive) prior on the
strength of the external shear, although this prior is not
particularly important.

Finally we place a Gaussian prior on the density slope of the
secondary galaxy (G2), for which we have no kinematic information.  We
choose $\gamma'_{\rm G2}=2.00\pm0.10$, fully consistent with the
spread in values of $\gamma'$ found from 0047--285 and MG2016+211
\citep{TK02a,KT03}.  Since G2 is $\sim$5 times less massive than G1,
its contribution to the properties of the lens system is less
pronounced. We discuss this prior further in \S7.3.

\subsection{Isothermal lens model}

To allow a comparison with previous modeling efforts
\citep{KF99,cskrings,F02}, we first set the density slopes of both G1
and G2 to isothermal (i.e.~$\gamma'$=2). The structure of the ring is
included as a constraint and the priors are as discussed in \S6.1.

We have listed the model parameters of the minimum-$\chi^2$ solution
in Table~4, where this model is designated as SIE. Compared with
previous isothermal lens models \citep{KF99,F02}, we find that the
inferred value of the Hubble Constant, H$_0$=71$^{+5}_{-6}$~\kmsmpc\
(68\% C.L), has increased by about 6--8~\kmsmpc. The error includes
all random errors and correlations between free parameters. This
increase is partly due to the two main lens galaxies being closer to
each other after extinction correction. Although each galaxy is
isothermal, their combined potential becomes steeper, leading to an
increase in H$_0$ \citep[e.g.,][]{wucknitz02, nonparam}. Other
contributions come from including external shear and widening the
errors on the image flux ratios and positions, which change the
$\chi^2$ surface.

\begin{table*}[t]
\caption{Recovered point-image properties from the minimum--$\chi^2$ SPLE1+D (isotropic) model.}
\centering
\begin{tabular}{crrrr}
\hline
\hline
 Image & $\Delta x$ ($''$) & $\Delta y$ ($''$) & $S_{\rm norm}$ & $\Delta t$ (days) \\ 
\hline
 A &   $+$0.00005  &  $+$0.00003 &  2.0408 & 32.42 \\
 B &   $-$0.73823  &  $-$1.96125 &  0.9313 &  0.00 \\   
 C &   $-$0.74461  &  $-$0.45379 &  1.0869 & 36.20 \\          
 D &   $+$1.12845  &  $-$1.25630 &  0.4230 & 76.51 \\          
\hline
\hline
\end{tabular}
\end{table*}

Besides reproducing all point-image constraints well, an excellent fit
(nearly identical to Fig.~5) to the boxy structure of the Einstein Ring
is also found. A comparison by eye to the fit in Kochanek et
al. (2001; their Fig.~5) shows that our fit is significantly better,
in particular between the image pairs B--C and D--A. The difference
can probably be attributed to the fact that we use a deconvolved
Einstein Ring shape and extinction-corrected lens galaxy centroids. We
find a source axial ratio of $q_s=0.64\pm0.03$ and position angle of
$\theta_s=(113\pm 2)^\circ$. The axial ratio is similar, but the
position angle of the ring source is significantly different from the
value in \citet{cskrings}. The optical source has a minor offset from
the source for the radio images ($\la$0\farcs03) which could, for
example, be attributed to a small asymmetry (e.g. lopsidedness or $m$=1
mode) of the host galaxy or deviations from ellipticity. Forcing the
centroid of the host galaxy to coincide with the radio source
increases the $\chi^2$ of the ring fit (by $\sim$30\%), but changes
the resulting mass model parameters only marginally. Since $m$=1
modes for the brightness distribution of galaxies are not uncommon and
automatically accounted for in models with a free host-galaxy
centroid, we choose not to force such a coincidence. We find values
identical to those derived from the F160W image when the ring shape
from the F814W data is used, and also when using the code from
\citet{lensmodel}.

As a further check, we calculate the stellar velocity dispersion of G1
based on this isothermal lens model. Details of the calculation method
are outlined in \citet{KT03} and \S4. We also come back to this in
more detail in \S6.4. We find $\sigma_{\rm ap}=230$~\kms\ for
isotropic models ($r_i=\infty$) and $\sigma_{\rm ap}=250$~\kms\ for
anisotropic models with $r_i=R_{\rm eff}$, inside the proper aperture
and corrected for seeing effects. These values agree remarkably well
with the measurement of $\sigma_{\rm ap}=247\pm 35$~\kms, even though
this information has not been used in the isothermal models. This
suggests that the simplest isothermal lens model is probably not too
far from the final model.

\subsection{Power-law lens models}

The next step is to allow the slope of the dominant lens galaxy (G1)
to vary. Since the isothermal models show that G1 is $\sim$5 times
more massive than G2, the contribution of G2 to the structure of the
Einstein Ring and the lensed images is relatively small. In general we
find that the properties of G2 are ill-defined if we leave its
density slope (or ellipticity; see \S7.3) entirely
unconstrained. We therefore place a prior on its slope, $\gamma'_{\rm
G2}=2.00\pm 0.10$, as suggested from similar mass models of other lens
systems (\S6.1). The slope of G1 is left fully unconstrained. We
designate this model as the ``Singular Power-Law Ellipsoid''
(SPLE) mass model. The SPLE1 model is that model with a
prior {\sl only} on the power-law slope of G2.

The results are listed in Table~4. We find a slightly better fit to
the data (i.e. $\Delta \chi^2$=$-1.8$) than for the isothermal models.
Most notably, the slope of G1 is found to be $\gamma'_{\rm
G1}=1.99^{+0.20}_{-0.20}$, nearly identical to isothermal. The slope
of G2 is somewhat steeper than isothermal, which leads to a slight
increase in the Hubble Constant, H$_0$=74$^{+10}_{-11}$~\kmsmpc\ (68\%
C.L) compared with the SIE mass models. Since the density slopes are
included as free parameters, the resulting error on H$_0$ is larger
than for the isothermal models which have fixed density slopes. The
error on H$_0$ is less than 15\%, even without kinematic
constraints. If we include a similar prior on the slope of G1 as on G2
(the SPLE2 model), the results are again nearly identical
(Table~4). The error on H$_0$ decreases considerably, because it is
dominated by the uncertainty on the density slopes, as discussed
previously. As for the isothermal models, the point-image constraints
and the shape of the Einstein Ring are reproduced well.

\subsection{Lensing \& Dynamical models}

The final step in the modeling effort is to include the stellar
velocity dispersion of G1 as an additional constraint. This is not as
simple as for a case with a single lens galaxy. Since we have two
lens galaxies, we can not simply attribute all of the mass within the
Einstein radius (or critical curve) to only the lens galaxy G1.

We therefore follow a different approach. First we determine the
minimum--$\chi^2$ parameters for a range of models with fixed values
of $\gamma'_{\rm G1}$ ranging from 1.6 to 2.4. We then calculate the
mass enclosed within an aperture with a radius of 1$''$ centered on
G1, using the lens model parameters. Using the enclosed mass and the
slope of the density profile of G1, we can then determine the stellar
velocity dispersion inside the observed aperture \citep[for details,
see \S 4 and][]{KT03}. The model velocity dispersion is compared to
the observed value and the resulting $\Delta \chi_{\sigma}^2$,
assuming a Gaussian error, is added to the $\chi_{\ell}^2$ from the
lens constraints alone.

Since $b$, $q$ and $\gamma'$ all enter into the determination of the
mass enclosed within 1$''$, the minimum of $\chi_{\rm l}^2 + \Delta
\chi_{\sigma}^2$ is not necessarily the true minimum. However, since
$\gamma'_{\rm G1}$ is by far the parameter that most dominates the
determination of the inferred stellar velocity dispersion\footnote{We
note that the effect of changing the aperture, effective radius,
etc. are completely negligible here.}, for all practical purposes the
difference between this approach and the correct approach (which is to
calculate the dispersion at every $\chi^2$ evaluation) is negligible
compared to the final error on H$_0$. We note that calculating the
dispersion at every minimization step would slow down the minimization
process by several orders of magnitude and is not feasible.

The resulting best-fit models for the lensing and dynamical models,
designated as SPLE1+D and SPLE2+D, are listed in Table~4.  We do
calculations for two different anisotropy radii, for $r_i=\infty$
(isotropic) and $r_i=R_{\rm eff}$ (anisotropic). Since the observed
stellar velocity dispersion is so close to that already predicted from
the lens models alone, we see that the final results barely differ
from the results presented in \S6.3, except that the error on H$_0$
has decreased. In particular we find that the SPLE2 models have nearly
identical errors on H$_0$ as the SPLE1+D models.  Hence, both the lens
and lens plus dynamics models clearly prefer an isothermal mass
distribution for G1.  In Table~5, we have listed the recovered
point-image properties from the best SPLE1+D model, assuming
$r_i=\infty$. Nearly identical results are found for $r_i=R_{\rm
eff}$. The best-fit model of the Einstein Ring is shown in Fig.~5,
along with the observed ring. Fig.~6 shows the critical and caustic
curves and the time-delay surface. For comparison, the outer critical curve
is also overplotted on the data on the left panel of Fig.~5. Note that
the critical curve is consistent with going through the saddle points
of the ring, as expected. This provides a consistency check of the
model, independent of the assumption of ellipticity for the source.

\section{The Hubble Constant from B{\footnotesize 1608+656}}

In this section we discuss the value of H$_0$ from B1608+656 and its
random and systematic errors in more detail, based on the class of
models presented in \S6.

\subsection{Models Using Only Lensing Constraints}

The SIE and SPLE1 and SPLE2 models are mass models based solely on the
gravitational lensing constraints (i.e., no stellar dynamics) and give
a value of H$_0$ ranging from 71--74~\kmsmpc\ (Table~4). We adopt the
value of H$_0$=74$^{+10}_{-11}$~\kmsmpc\ as the best lensing-only
determination, which includes the uncertainty on the density slope of
the G1 without any assumptions except for a prior on the slope of
G2. We discuss the latter in more detail in \S7.3. We note
that including a prior on G1 similar to that on G2 does not change the
result, since the slope of G1 is already found to be nearly
isothermal, i.e., $\gamma_{\rm G1}'=1.99^{+0.20}_{-0.20}$ (68\% CL)
without a prior. The lensing-only models therefore predict an
isothermal density profile for G1 to within~10\%.

\subsection{Models Using Lensing and Dynamical Constraints}

All other models we consider include the stellar velocity dispersion
of G1 as a constraint. In \S6, we showed that the simplest lens models
that we consider, with two isothermal galaxies, predict a stellar
velocity dispersion close to the observed value (within 1--$\sigma$
for both the isotropic and strongly anisotropic models).  This
agreement shows that the slope of the density profile of G1 can not
deviate from isothermal too strongly. Indeed it is found that the
slope of G1 is isothermal to within a few percent if we include the
stellar velocity dispersion as an additional constraint.  We finally
adopt the median value of H$_0$=75$^{+7}_{-6}$~\kmsmpc\, as the best
determination. A systematic error of $\pm$\,1~\kmsmpc\, is estimated,
due to the unknown anisotropy of G1. We note that including the
stellar dynamics has not significantly changed the value of H$_0$, but
reduces its error considerably, to less than 10\%. Similarly, the
slope of G1 is $\gamma_{\rm G1}'=2.03^{+0.14}_{-0.14} \pm 0.03$ (68\%
CL), where we take the average value of the SPLE1+D models with
$r_i=R_{\rm eff}$ and $r_i=\infty$. The latter error is the
uncertainty due to the unknown anisotropy of the stellar velocity
ellipsoid of G1 (Table~4).

\subsection{Systematics}

Finally we discuss the dominant systematic errors that could
potentially be left in the models.

The results have not drastically changed since earlier models, but
given the wealth of new constraints (e.g., the Einstein Ring, improved
time delay measurements, stellar dynamics, etc.), the results have
become more robust.  The error on H$_0$ has shrunk even
though we have included more free parameters (e.g., external shear,
varying slopes and galaxy positions, etc.) in the model. The smaller
error on H$_0$ is due to improvements in the measurement of the time
delays and the inclusion of the Einstein Ring and stellar
dynamics. Support for the refined models is further given by the flux
ratio of image D, which is recovered to within 22\% (Tables~3 and 5),
even though it has not been included as a constraint on the
models. All previous models had great difficulty with this. Hence, at
first glance, the results appear little affected by remaining
systematics.

Even so, we remain careful about assumptions, the most important of
which is that for the prior on the density slope of G2 (\S6.1). If we
increase the width of the prior, we find that H$_0$ increases somewhat
(i.e., the density profile of G2 becomes steeper), but also that its
axial ratio decreases to very small values $\la 0.2$ (PA remains
similar to those in Table~4). Since G2 is most likely a disk galaxy
(because of the copious amounts of dust associated with it), a small
axial ratio is not unlikely, but a value of $\sim$0.1--0.2 appears
unrealistic, especially since G2 has no discernible ellipticity in the
F160W image (Fig.~1). If we take priors of $q_{\rm G2}=$1.0$\pm$0.2 and
loosen the prior on the density slope to $\gamma'_{\rm
G2}$=2.0$\pm$0.3 -- both of which are also reasonable, given the
observations -- we find that the resulting value of H$_0$ increases by
at most a few percent.

\begin{figure*}[t]
\begin{center}
  \leavevmode
\hbox{%
  \epsfxsize=\textwidth
  \epsffile{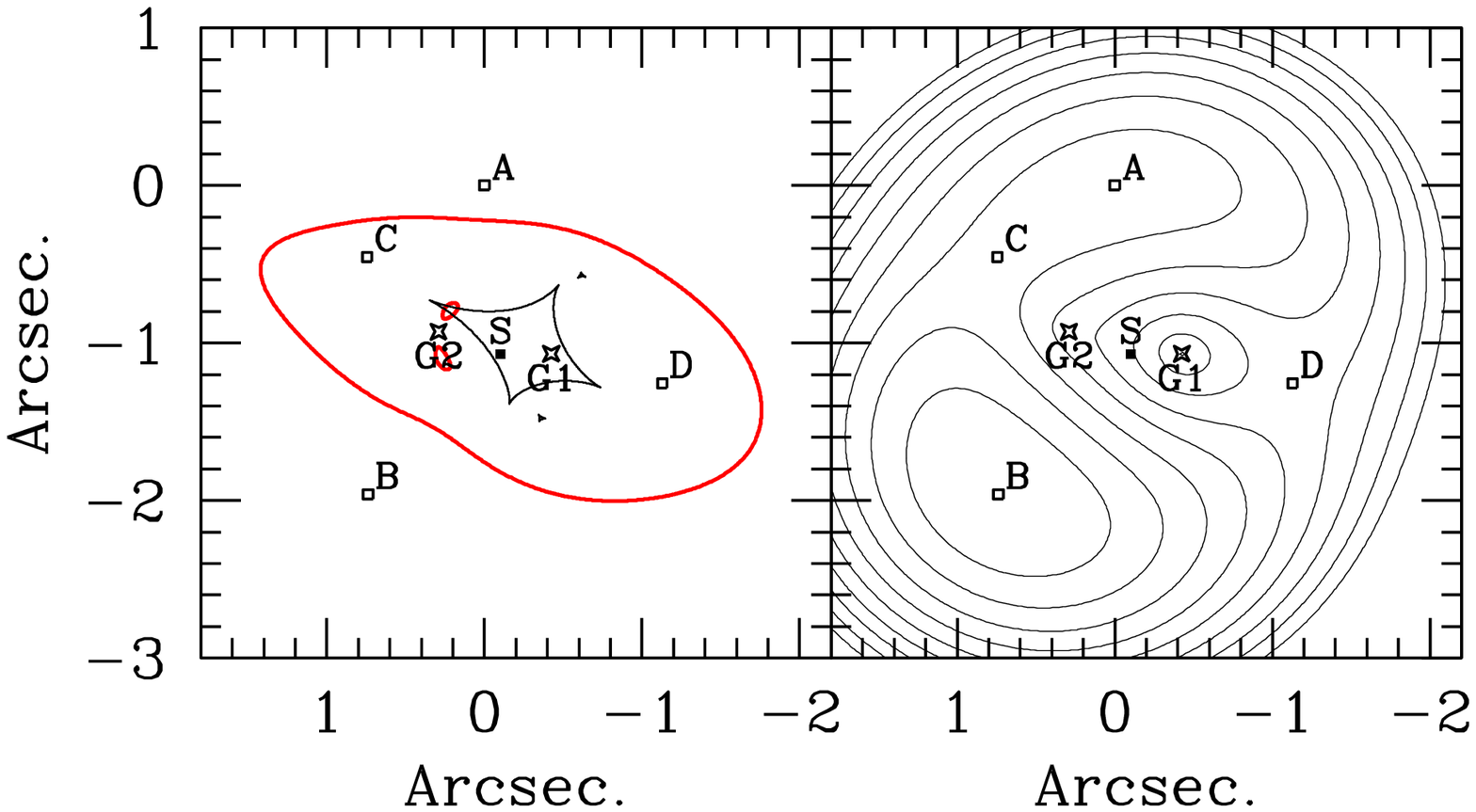}}
\end{center}
\caption{Left: Critical (thick) and caustic curves (thin) of the
SPLE1+D model. The galaxy positions are indicated by stars, the images
positions by open squares and the source by a closed square. Right:
The contours indicate constant time delays starting at $\Delta t$=0 at
image B and increasing in steps of 10 $h^{-1}$ days.}
\end{figure*}

Next, we have assumed a spherical mass distribution for G1 in our
stellar dynamical models. Since we measure a luminosity-weighted
dispersion within a relatively large aperture ($\ga R_{\rm eff}$),
many assumptions about the phase-space density of stars have little
effect on the final measured stellar velocity dispersion. For example,
anisotropy has little effect on H$_0$. A change of $\pm$5\% in the
average dispersion would also affect H$_0$ on a level of $\la$3\%. We
therefore believe the assumption of spherical symmetry to be of little
effect, given the current error on the observed stellar velocity
dispersion of G1 ($\sim$14\%). We also notice that the slope of G1
from lensing alone and from stellar dynamics fully agree, giving no
indication of large remaining systematic errors.

As a check on the astrometry, dust correction and deconvolution
procedure, we have used the Einstein Ring from the F814W observations
as a constraint. We follow exactly the same procedure as for the F160W
image. We find that the two lens galaxies lie slightly further apart
than in the NICMOS images, as discussed previously. This suggests that
a smaller value of H$_0$ will be found.  However, the best SPLE1+D
model produces a somewhat steeper density profile for G1 (by 2\%),
which compensates for the somewhat larger lens galaxy separation and
leads to an identical value of H$_0$. Hence, it appears that inclusion
of the ring ``forces'' the combined lens potential of G1 and G2 to be
slightly steeper than isothermal, leading to a robust value of H$_0$
independent of which data set is used.

We therefore conclude that given the assumptions and priors -- which
are all consistent with the observations of B1608+656 -- remaining
systematic errors on H$_0$ probably add up to not more than about 5\%,
dominated by those from assumptions in the stellar dynamical model and
the priors on G2.

However, an important systematic error that could still affect the
value of H$_0$ is a mass-sheet degeneracy \citep{masssheet} due to the
small group around the lens system \citep{2002AAS...201.8009F}.
Although our model reproduces the observed stellar velocity
dispersion well for an isothermal mass distribution, a strong
mass-sheet could have led us to overestimate the enclosed galactic
mass and therefore to underestimate the inferred density slope, given
the observed velocity dispersion constraint. The observed
quasi-isothermality, however, does not necessarily argue against a
mass-sheet since the mass density profile does not have to be exactly
isothermal. We note also that the effect on the slope partially
compensates the well known scaling of H$_0$ with $\kappa_c$ for a
given mass model. We defer a precise quantification of the convergence
($\kappa_c$) and shear of this group, and their effects on the lens
system and H$_0$, to a forthcoming publication with additional data on
the group (Fassnacht et al.~in~prep.), but believe it to be relatively
small based on the ``poorness'' of the group and its small velocity
dispersion.

Finally, we note that the effects of the cosmological model are
negligible. Changing $\Omega_{\Lambda}$ from 0.7 to 0.0, with
$\Omega_{\rm m}=0.3$, increases H$_0$ by $2\%$. A drastic change to
$(\Omega_{\rm m}=1.0, \Omega_{\Lambda}=0.0)$ decreases H$_0$ by
$7\%$. However, within the current limits on the cosmological model set by
WMAP\citep{spergel03}, changes in H$_0$ are $\la 1\%$.

\section{Discussion \& Conclusions}

We have presented significantly improved and refined mass models of
the gravitational lens B1608+656 -- compared with previous modeling
efforts -- with the aim of determining a value of the Hubble Constant,
that is less affected by previously known systematics (e.g. radial mass
profile, dust extinction, etc.).

New constraints on the mass model include: (i) the stellar velocity
dispersion of the dominant lens galaxy (G1), as measured with ESI,
(ii) the deconvolved Einstein Ring seen in the HST F160W and F814W
images -- the former of which is little affected by dust -- corrected
for the contribution from the lens galaxies, (iii) the
extinction-corrected lens-galaxy centroids and structural parameters,
the former being one of the major uncertainties in previous lens
models and (iv) recent improvements in the determination of the three
independent time delays in this four-image lens system \citep{F02}.

Lens models have also been improved in allowing for many additional
free parameters compared with previous modeling efforts
\citep{KF99,F02}, including the galaxy positions, the position angle
of G1, an external shear and the density slopes of G1 and G2. Some of
these parameters are constrained with observational priors. The
freedom in the lens model (up to 22 free parameters) allows for a
proper analysis of the error on the inferred value of H$_0$, including
all observational errors and correlations between free parameters.

The improvements in the observations {\sl and} lens models have led to
a Hubble Constant from B1608+656 which is less affected by known
systematic errors than previously.  We obtain
H$_0$=75$^{+7}_{-6}$~\kmsmpc\ (68\% CL), with a $\pm$5\% error
contributed by known systematic errors (for $\Omega_{\rm m}$=0.3 and
$\Omega_{\Lambda}$=0.7). This value is higher than found from
previous models \citep[e.g.,][]{KF99,F02}, predominantly because,
after extinction correction, the two lens galaxies are somewhat closer
together than previously thought, their density profiles are slightly
steeper than isothermal and external shear is included in the models.

First, we find that the new lens models reproduce all radio-image
constraints to within the errors. This agreement might imply that we
have been too conservative in the error estimates and might also have
slightly overestimated the final error on H$_0$. The flux ratio of
image D is also reproduced to within 22\%, a considerable improvement
over previous models \citep[e.g.,][]{KF99,F02}. Image D is the closest
image to the dominant lens galaxy and might be affected
by structure in the stellar (or dark-matter) mass distribution of lens
galaxy G1 that is poorly represented by the global mass model.

Second, the lens models reproduce the observed stellar velocity
dispersion of G1, and its mass profile is found to be nearly
isothermal, $\gamma_{\rm G1}'=2.03^{+0.14}_{-0.14} \pm 0.03$ (68\%
CL).  This result is not as straightforward as one might expect, since
the inferred stellar velocity dispersion of G1 depends strongly on the
mass that G2 contributes to the lens properties. If the lens model
predicted too large a mass for G2, the mass of G1 would decrease
(because lensing tightly fixes the mass enclosed by the lensed
images), as would the inferred stellar velocity
dispersions. Similarly, if the mass is correct, but the density slope
of G1 inferred from lensing alone is wrong, the inferred velocity
dispersion would be incorrect as well. The fact that the lensing model
(without dynamical constraints) already predicts a mass model for G1
that accurately reproduces the observed stellar velocity dispersions
makes the models, in our opinion, very believable. The addition of the
dynamical constraint therefore does not change the results from
lensing alone, but significantly tightens the error on the value
H$_0$.

Third, the lens models reproduce the boxy shape of the Einstein Ring
both in the F160W and F814W bands well and give the same value of
H$_0$ irrespective of which of these datasets we use as a constraint.
In addition, the critical curves (see Fig.5) cross the Einstein Ring
at the saddle points of its brightness distribution, as required
\citep[e.g.,][]{rdb1608}.

We note that the resulting value H$_0$ from B1608+656 seems difficult
to reconcile with \citet{2002ApJ...578...25K, 2003ApJ...583...49K},
who finds that perfectly isothermal lens galaxies should lead on average
to H$_0=48\pm3$\,\kmsmpc\ (see also Kochanek \& Schechter 2003). We
find a significantly higher value of H$_0$ even if the dominant lens
galaxy (G1) is {\it assumed to be} isothermal. One possible
explanation is that the time-delay systems examined by
\citet{2002ApJ...578...25K, 2003ApJ...583...49K} have lens galaxies
with luminous plus dark-matter mass profiles steeper than isothermal,
enough to yield H$_0$ in better agreement with that obtained from
B1608+656. Although indeed -- {\it on average} -- isothermal models
appear to provide a good representation of the mass distribution of
early-type galaxies \citep[e.g.][Rusin et al.\ 2003b]{TK02a,KT03},
significant departures in several individual cases is not necessarily
inconsistent with the observed spread of mass slopes of local
early-type galaxies (e.g., Fig.~1 in Gerhard et al.\ 2001; Bertin \&
Stiavelli 1993), especially at small radii where the stellar mass
starts to dominate (e.g. Bertola et al.\ 1993). The lens galaxy in
PG1115+080 \citep{TK02b} provides one such example of a lens-galaxy
mass profile steeper than isothermal inside the Einstein radius.

Conversely, a mass-sheet degeneracy could have led us to overestimate
H$_0$ from B1608+656, although the mass-sheet degeneracy is somewhat
broken by the kinematic measurement. In this respect, we only note
that a convergence of at least $\kappa_c$$\sim$0.4 is needed to bring
our value of H$_0$ in agreement with $48\pm3$\kmsmpc, an
unrealistically high value, and we defer further discussion to a
forthcoming paper (Fassnacht et al.\ in prep.).

Another viable alternative would be that the time-delay systems
examined by \citet{2002ApJ...578...25K, 2003ApJ...583...49K} have not
been modeled in enough detail yet and could also be affected by nearby
clusters or groups as seen near some of these systems \citep[e.g.,
RXJ0911;][]{2000ApJ...544L..35K, 2002ApJ...572L..11H}.  A simple
prediction is therefore the following: {\sl If} these lens systems are
isothermal and {\sl not} affected by external perturbers, then the
stellar velocity dispersion should be that predicted from a simple
isothermal lens model (as for B1608+656). {\sl If}, on the other hand,
the observed stellar velocity dispersion is much higher that predicted
from isothermal models, this would be a clear sign of a steeper mass
profile (as for PG1115+080). Hence, stellar kinematics of lens
galaxies is a key measurement to further study this apparent
inconsistency.

Even so, the analysis of B1608+656 has demonstrated that when enough
information is available, it is possible to reach accuracies of
$\la$10\% (random) on the value of H$_0$ (or $\la$15\% when including
systematic errors) taking the radial mass profile properly into
account. A similar uncertainty was obtained by applying a joint
lensing and dynamical analysis to PG1115+080 \citep{TK02b}. This level
of uncertainty is comparable to the accuracy achieved by the
individual methods (SNae--Ia, Fundamental Plane, etc) that contributed
to the final value of H$_0$ from the HST Key-Project
\citep{keyproject}. Hence, obtaining sufficiently good data -- including
on the local environments -- for a handful of gravitational-lens
systems, seems to be a promising way to break the 10\% limit on the
uncertainty on the value of H$_0$.

Finally, the analyses of B1608+656 and PG1115+080 have demonstrated
that there are no simple ways to reach this goal and it does not
suffice to assume a radial mass profile, without measuring it for each
individual system. That approach probably does not lead to a reliable
value of H$_0$. High-quality observations and detailed lens mass
models -- that allow for many degrees of freedom -- are required to
obtain an accurate (and realistic) estimate of the Hubble Constant and
its uncertainty. We are currently collecting spectroscopic data for
additional lens systems with measured time-delays, with this aim in
mind. In addition, optical and radio monitoring programs are
continuing that should provide more (and/or improved) time delays in
the near future, and forthcoming observations with HST could provide
improved constraints on the lens galaxies.

{\acknowledgments We thank Richard Ellis and Jean-Paul Kneib for
useful comments on this manuscript and stimulating conversations. We
are grateful to an anonymous referee for comments that helped
clarifying the manuscript. The use of the Gauss-Hermite Pixel Fitting
Software developed by R.~P.~van der Marel is gratefully
acknowledged. The ESI data were reduced using software developed in
collaboration with D.~Sand.  We acknowledge the use of the HST data
collected by the CASTLES collaboration. LVEK and TT acknowledge
support by NASA through grant AR-09960 from the Space Telescope
Science Institute, which is operated by the Assocation of Universities
for Research in Astronomy, under NASA contract NAS5-26555. LVEK also
acknowledges a STScI Fellowship grant. LVEK and RDB acknowledge NSF
grants AST 99-00866 and 0206286, and GS acknowledges a NASA grant NAG
5--7007 and CONICET in Argentina. We thank J. Miller, M. Bolte,
R. Guhathakurta, D. Zaritsky and all the people who worked to make ESI
such a nice instrument. Finally, the authors wish to recognize and
acknowledge the very significant cultural role and reverence that the
summit of Mauna Kea has always had within the indigenous Hawaiian
community.  We are most fortunate to have the opportunity to conduct
observations from this mountain.  }

\clearpage

\clearpage

\clearpage

\clearpage

\clearpage

\end{document}